 \documentclass[aps,prb,reprint,twocolumn]{revtex4-1}
 \usepackage{amsmath}
 \usepackage{amssymb}
 \usepackage{graphicx}
 \usepackage{bm}
 \usepackage{color}

\begin{document}

\title{Topological Edge States Induced by Zak's Phase in A$_{3}$B
  Monolayers}
\author{Tomoaki Kameda$^1$, Feng Liu$^1$, Sudipta Dutta$^2$, Katsunori Wakabayashi$^{1,3}$}
\email[Corresponding to: ]{waka@kwansei.ac.jp}
\affiliation{$^1$Department of Nanotechnology for Sustainable Energy,
  School of Science and Technology, Kwansei Gakuin University, Gakuen 2-1, Sanda,
  Hyogo 669-1337, Japan}
\affiliation{$^2$Department of Physics, Indian Ins
  titute of Science
Education and Research (IISER) Tirupati, Tirupati 517507, Andhra
Pradesh, India}
\affiliation{$^3$National Institute for Materials Science (NIMS), Tsukuba, Ibaraki 305-0044, Japan}

\begin{abstract}
In crystalline systems,
charge polarization is related to Zak's phase determined
by bulk band topology. Nontrivial charge polarization induces robust
edge states accompanied with fractional charge.
In Su-Schrieffer-Heeger (SSH) model,
it is known that the strong modulation of electron hopping causes
nontrivial charge polarization
even in the presence of inversion symmetry.
Here, we consider a bi-atomic honeycomb lattice to
introduce such strong modulation, \textit{i.e.} A$_3$B sheet.
By tuning hopping ratio and onsite potential difference between A and B
 atoms, we show that topological
phase transition characterized by Zak's phase occurs.
Furthermore, we propose that C$_3$N and BC$_3$ are
the possible realistic materials on the basis of first-principles calculations.
Both of them display topological edge states induced by Zak's phase without
spin-orbital couplings and external fields unlike conventional
 topological insulators.
\end{abstract}

\maketitle
\section{Introduction}
Concept of topology~\cite{Bansil2016} leads to a new class of electronic materials, such as topological
insulators,~\cite{Kane2005,Bernevig2006,Hsieh2008,Chen2009,Chang2013,Ando2013,Sato2016}
topological crystalline
insulators,~\cite{Fu2011,Tanaka2012,Dziawa2012a} and Weyl
semimetals.~\cite{Wan2011,Borisenko2014,Lu2015}
In topological materials, topologically protected edge states (TES) emerge
owing to nontrivial bulk band topology. These TES are
robust to defects and edge roughness,
and can be exploited for applications to low-power-consumption
electronic and spintronic devices.~\cite{Checkelsky2012}
One origin of TES is nonzero Berry curvature introduced by
spin-orbit couplings.
Berry curvature is a geometric field strength in momentum
space. Its integration over momentum space yields magnetic monopole
that is characterized by Chern number.~\cite{Xiao2010,Haldane2006}

Recently two of us have found that, even under zero Berry curvature, Berry connection -- a geometric vector
potential whose curl yields Berry curvature -- can also lead to TES.~\cite{Liu2017,Liu2017B,Liu2018}
Integration of Berry connection over momentum space (also called as Zak's phase)~\cite{Zak1989,
Delplace2011} results in an electric dipole moment that generates robust
fractional surface charges.~\cite{King1993,Resta1994,Zhou2015}
Such dipole field related to Zak's phase brings a new type of
topological materials, \textit{i.e.} topological electrides.~\cite{Huang2018,Hirayama2018}

To obtain nonzero Zak's phase even in the presence of inversion
symmetry, modulation of electron hopping is
necessary. Employing Su-Schrieffer-Heeger (SSH)
model~\cite{Heeger1988} on two-dimensional (2D) square lattice,~\cite{Liu2017}
nontrivial Zak's phase emerges when the inter-cellular hopping is
larger than the intra-cellular hopping. This can be successfully
demonstrated in the photonic system by mimicking the electronic
tight-binding model.~\cite{Liu2018}
In addition, this idea can also be extended to honeycomb lattice
systems with Kelul\'e pattern.~\cite{Liu2017B}
However, no realistic materials of nonzero Zak's phase
have been proposed in this framework yet. Also, the model proposed in
Ref.~\onlinecite{Liu2017B} is hard to apply for designing atomistic model,
since it demands strong hopping modulation in monatomic sheets.

To overcome this difficulty, we consider biatomic system of
honeycomb lattice, \textit{i.e.} A$_3$B atomic sheet.
We show that TES emerge owing to different
electron hopping and onsite potentials between A and B atoms.
Furthermore, we propose two possible realistic materials, \textit{i.e.} C$_3$N and
BC$_3$ based on first-principles calculations.
Both C$_3$N and BC$_3$ display TES induced by Zak's phase.
Remarkably, both BC$_3$ and C$_3$N have already been successfully
synthesized by several experiments.~\cite{Yanagisawa2004,Tanaka2006,Mahmood2016}


The paper is organized as follows.
In Sec. II, we relate charge polarization to Zak's phase in terms
of Berry connection in 2D crystalline systems. We especially discuss the cases where energy
bands are degenerate.
In Sec. III, we investigate electronic states of A$_3$B monolayer and
their zigzag nanoribbon (NR) on the basis of tight-binding model.
We show the bulk-boundary correspondence, \textit{i.e.} emergence of
TES and nonzero Zak's phases.
In Sec. IV, we analyze the electronic structures of
C$_3$N and BC$_3$ sheets and their nanoribbons on the basis of
first-principles calculations. We verify the existence of TES in energy bands for $\pi$ electrons in these materials.
We summarize our results in Sec.V.

\begin{figure*}[ht]
  \begin{center}
    \includegraphics[width=0.95\textwidth]{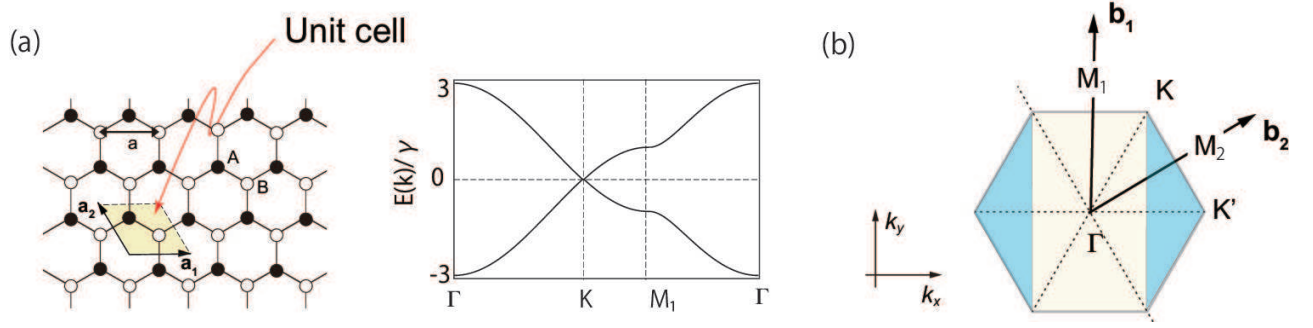}
  \end{center}
\caption{
(a) Lattice structure of graphene.
Shaded yellow rhombus indicates unit cell, which contains
sublattices A and B.
$\bm{a_1}=(a,0)$ and $\bm{a_2}=(-\frac{a}{2},\frac{\sqrt{3}a}{2})$ are
primitive lattice vectors. Inset displays energy band structure of
graphene where energy are degenerate at $K$ and $K^\prime$ points.
(b) Corresponding first BZ of graphene.
$\bm{b_1}=\frac{2\pi}{\sqrt{3}a}(0,2)$ and
$\bm{b_2}=\frac{2\pi}{\sqrt{3}a}(\sqrt{3},1)$ are primitive reciprocal
vectors. Yellow and blue regions indicate valid and invalid areas for applying Eqs.~(3) and (4), respectively,
if $\bm{b_1}$ is chosen in the direction of charge polarization, \textit{i.e.} $P_1(k)$.
}
 \label{fig:graphene}
\end{figure*}
\section{Charge Polarization and Zak's phase}
Charge polarization can be regarded as geometric center of electronic
wavefunctions.
For 1D crystalline system, the charge polarization of $n$-th energy band
is given as a Wannier center, {\it i.e.}  $P^n=\langle w_n(x)|x|w_n(x)\rangle$,
where $w_n(x)$ is a Wannier function of $n$-th energy band.~\cite{Marzari2012} Owing to gauge freedom of Wannier
functions, $P^n$ is well-defined up to a lattice constant.

To relate $P^n$ to Berry connection, one can apply Fourier
transformation to $w_n$ and $x$, which result in~\cite{Marzari2012}
\begin{equation}
  P^n=\dfrac{1}{2\pi}\int_0^{2\pi}\langle\psi_n|i\partial k|\psi_n\rangle dk,
\end{equation}
where $\psi_n$ is the periodic part of Bloch function of $n$-th energy
band, and $A^n=\langle\psi_n|i\partial k|\psi_n\rangle$ is Berry
connection. We refer to the integration part of Eq.~(1) as Zak's phase.~\cite{Zak1989}
In a finite chain, fractional charge $P=\pm\sum_n^{n_{occ}}P^n$ accumulate at the ends
of the chain, where the summation is taken over all the occupied
energy bands.~\cite{Rhim2017} When inversion symmetry is present, $P^n$ is quantized to
0 or $1/2$, and is determined by the winding number of associated
sewing matrix of wavefunctions.~\cite{Fang2012a}
In following discussions, we interchangeably use ``Zak's phase'' and
``charge polarization'' to mean same quantity.

Now we extend Eq.~(1) to 2D crystalline systems. Suppose that there are two
independent directions denoted as $i$ and $j$. In 2D systems, charge polarization
becomes a vector such as $\mathbf{P}^n=(P^n_i,P^n_j)$, and $P^n_i$, $P^n_j$ depend
on the wavenumbers $k_j (=k)$, $k_i$ along directions $j$, $i$, respectively.
Charge polarization along $i$-direction is given as
\begin{equation}
 P^n_i(k)=\dfrac{1}{|\mathcal{P}|}\int_{\mathcal{P}} \mathbf{A}^n(k_i,k_j)
  \cdot \mathbf{n}_idk_i,
\end{equation}
where $\mathcal{P}$ is a straight path connecting two equivalent
$\mathbf{k}$ points in momentum space, $\mathbf{n}_i$ is a unit vector for
$i$-direction and
$\mathbf{A}^n=\langle\psi_n|i\partial \mathbf{k}|\psi_n\rangle$ is Berry
connection of $n$-th energy band in 2D momentum space.
Similar to 1D systems, fractional charge $\sum_n^{n_{occ}}\mathbf{P}^n\cdot \mathbf{n}_i$ accumulates
on the edge if a material possesses finite charge polarization.
The derivation of Eq.~(2) is given in supplement of
Ref.~\onlinecite{Liu2018B}.

When inversion symmetry is present, $P^n_i(k=0)$ is simply determined by
the parities at inversion-invariant points in BZ. In hexagonal
lattice, polarization along
$i$-direction at $k=0$ is given as~\cite{Fang2012a,Liu2017B}
\begin{equation}
  \label{eq:parity}
  P^n_i(k=0)=\dfrac{1}{2}(q^n_i \text{ modulo } 2), \text{    } (-1)^{q^n_i}=\dfrac{\eta^n(M_i)}{\eta^n(\Gamma)},
\end{equation}
where $i=1,2$, and $\eta^n(\mathbf{k})$ is the eigenvalue of $\pi$ rotation along the
out-of-plane direction for $n$-th energy band.

According to Stoke's theorem, the relation of $P^n_i(k_a)$ and $P^n_i(k_b)$
is given by~\cite{Resta1994}
\begin{equation}
  \label{eq:deltap}
  \Delta P^{n,i}_{k_a,k_b} \equiv P^n_i(k_a)-P^n_i(k_b)=\dfrac{1}{2\pi i}\iint \mathcal{F}^n
  (k,k^\prime) dk dk^\prime,
\end{equation}
where $F^n=\partial_iA_j-\partial_jA_i$ is Berry curvature of $n$-th
energy band, and the integration is taken over the area $|(k_a-k_b) \times \mathcal{P}|$.

Combining Eqs.~(\ref{eq:parity}) and
(\ref{eq:deltap}), one can obtain $P^n_i(k)$ at arbitrary $k$ from the parities and
Berry curvature of wavefunction of $n$-th energy band.
Depending on the positions where
Berry curvature is finite, the value distribution of $P^n_i$ over BZ
can be dissociated into several distinct areas.
Note that to apply Eqs.~(4), a gap-opening condition
must be satisfied, {\it i.e.} $|E_n(k)-E_{n\pm 1}(k)|\neq 0$.

In degenerate systems, both Berry connection and Berry curvature
are written in the non-abelian forms. In this situation, 
we need to use more generic gap-opening condition:
$E_n(\mathbf{k})\neq E_{n^\prime}(\mathbf{k})$ for all $k$, where
$n\in I$ and $n^\prime \notin I$ for
$I=\{n_1,n_2,\cdots,n_{occ}\}$.~\cite{Fukui2005}
However, when we consider Zak's phase exactly at degenerate $k$ point,
even the generic gap-opening condition cannot be satisfied.
In such the cases, we apply an
inversion-symmetry-preserving perturbation to lift those
degeneracies, as far as the perturbation does not alter the order of
parities of energy bands at inversion-invariant $\mathbf{k}$ point.
If such the generic gap-opening condition cannot be satisfied by imposing the
perturbation, Eqs.~(3) and (4) can be applied only for the region without degeneracies.

Let us take graphene as an example of degenerate systems. Figure 1(a) displays graphene lattice structure and its bulk energy
bands, and Fig.~1(b) displays its corresponding BZ where yellow and blue
areas indicate valid and invalid ranges of applying Eqs.~(3) and (4),
respectively. In graphene, energy bands are degenerate at $K$ and
$K^\prime$ points guaranteed by inversion symmetry, and Eqs.~(3) and (4) can only be
applied in the range $-2\pi/3<k<2\pi/3$, if 
$\bm{b_1}$ is chosen in the direction of charge polarization, {\it i.e.} $P_1(k)$. Outside
this range, only Eq.~(2) is applicable.~\cite{Delplace2011}

\begin{figure*}[ht]
  \begin{center}
    \includegraphics[width=0.95\textwidth]{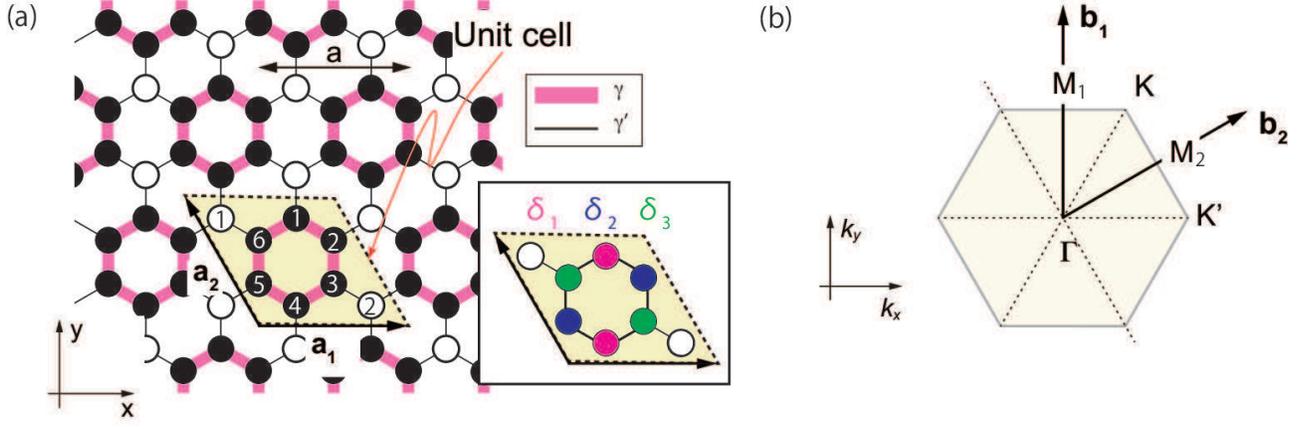}
  \end{center}
\caption{
(a) Lattice structure of A$_3$B biatomic sheet.
Shaded yellow rhombus indicates unit cell, which contains
six A-atoms (black circles indexed as $1,\cdots, 6$) and two B-atoms
(white circles indexed as 1 and 2).
$\bm{a_1}=(a,0)$ and $\bm{a_2}=(-\frac{a}{2},\frac{\sqrt{3}a}{2})$ are
primitive lattice vectors. Inset displays non-uniform onsite
potentials $\sigma_m \sim m\gamma/100$ ($m=1,2,3$) on A atoms, which breaks $C_{6}$ point
group symmetry and preserves inversion symmetry.
(b) Corresponding first BZ of A$_3$B atomic sheet.
$\bm{b_1}=\frac{2\pi}{\sqrt{3}a}(0,2)$ and
$\bm{b_2}=\frac{2\pi}{\sqrt{3}a}(\sqrt{3},1)$ are primitive reciprocal
vectors.
}
 \label{fig:A3B_structure}
\end{figure*}



\section{Tight-Binding Model}
\label{sec:tbm}

Let us introduce a tight-binding model for $\pi$ electrons up to nearest-neighbor hopping on a biatomic honeycomb
lattice A$_3$B. We show that TES appear
due to nonzero Zak's phase in A$_3$B atomic sheet.
The lattice structure of A$_3$B is displayed in Fig.~\ref{fig:A3B_structure}(a),
whose unit cell is a rhombus made up by six A-atoms (black circles indexed from 1
to 6) and two B-atoms (white circles indexed as 1 and 2). We denote
hopping between A-A (A-B) atoms as $-\gamma$ ($-\gamma^\prime$), and
onsite potential of A (B) atoms as $V_A$ ($V_B$). Figure~\ref{fig:A3B_structure}(b) shows the corresponding
first Brillouin zone (BZ), where the reciprocal lattice vectors are
$\bm{b_1} =\frac{2\pi}{\sqrt{3}a}(0,2)$ and $\bm{b_2} =\frac{2\pi}{\sqrt{3}a}(\sqrt{3},1)$.

The tight-binding Hamiltonian of A$_3$B can be written as
\begin{equation}
\hat{H}  =
\sum_m\sum_{\langle j,j^\prime \rangle}\sum_\alpha f^{\alpha}(j,j^\prime)\alpha^\dagger_{mj^\prime}\alpha_{mj}
-\gamma^\prime\sum_{\langle mj^\prime, nj \rangle} a_{mj^\prime}^\dagger b_{nj}
+{\rm H.c.},
\label{eq:hamiltonian}
\end{equation}
where $m, n$ are unit cell indices, $j$, $j^\prime$
($j, j^\prime=1,2\cdots6$ for A atoms and $j,j^\prime=1,2$ for B atoms)
are indices of atomic orbitals
in each unit cell, and $\alpha=a,b$.
${a^\dagger}$ ($b^\dagger$) and $a$ ($b$)
mean creation and annihilation operators of $p_z$ electronic orbital on
atom A (B), respectively.
${\langle\cdots\rangle}$ indicates the summation between the
nearest-neighbor sites. Here
\begin{equation}
  f^{\alpha}(j,j^\prime)=
  \begin{cases}
    -\gamma &\text{if $j\neq j^\prime$ and $\alpha=a$},\\
    V_A &\text{if $j=j^\prime$ and $\alpha=a$},\\
    V_B &\text{if $j=j^\prime$ and $\alpha=b$}.
    \end{cases}
\end{equation}
Note that, the electronic states of graphene recover when
 $\gamma=\gamma^\prime\approx 3$eV and $V_{A}=V_{B}=0$.

 \begin{figure*}[ht]
\includegraphics[width=0.9\textwidth]{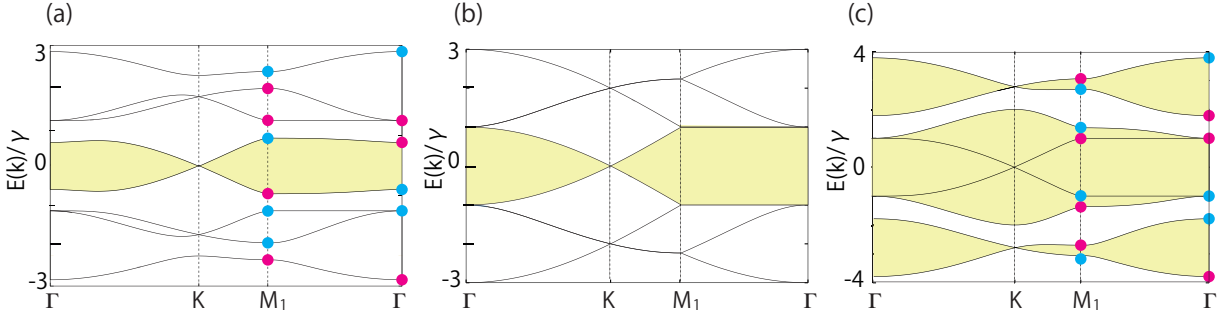}
\caption{Energy band structures of 2D A$_3$B sheet with
(a)~$\gamma^\prime=\frac{\gamma}{1.5}$,
(b)~$\gamma^\prime=\gamma$, and
(c)~$\gamma^\prime=1.5\gamma$. 
Onsite potentials are considered to be zero, {\it i.e.}
$V_{A}=V_{B}=0$.
Red (blue) circles at $\Gamma$ and $M_1$ points indicate the positive
(negative) parity of wavefunction.
Note that all three nonequivalent $M$ points have the same parity.
The yellow regions indicate the range where Zak's phase are
nonzero.
}
\label{fig:2Dband}
\end{figure*}
\begin{figure*}[ht]
\begin{center}
 \includegraphics[width=0.9\textwidth]{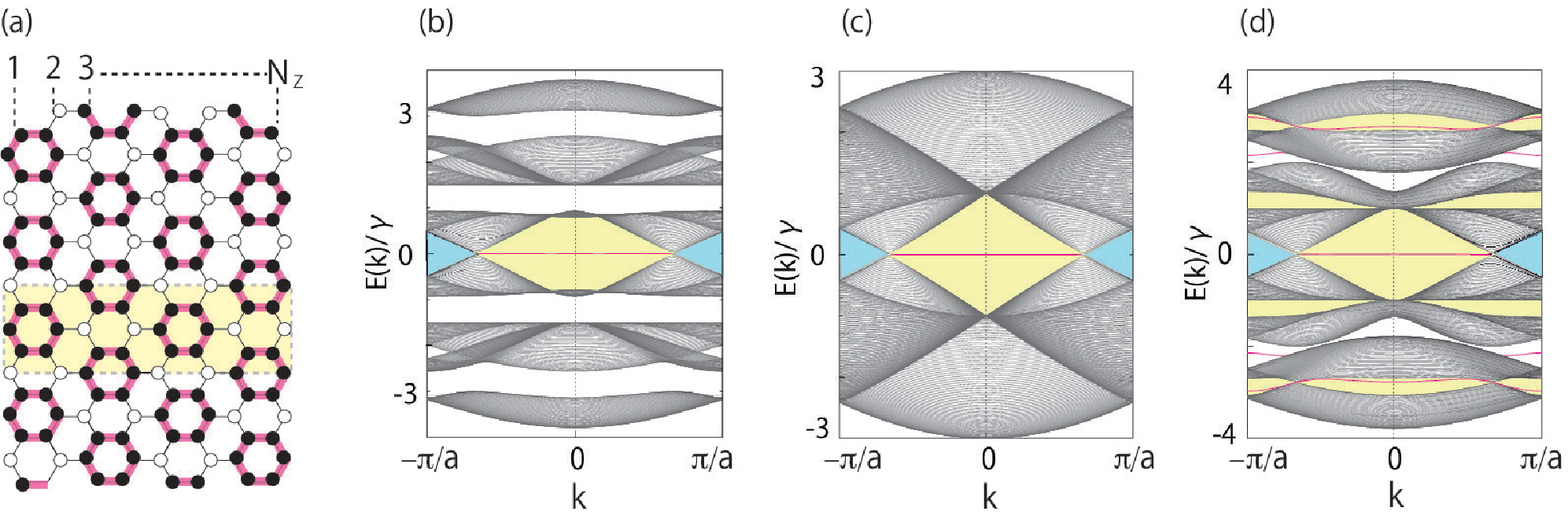}
\end{center}
\caption{(a)Lattice structure of A$_3$B zigzag NR.
Thick and thin bonds represent the intra-cell ($\gamma$) and inter-cellular ($\gamma^\prime$)
hopping, respectively. The yellow rectangle indicates unit cell.
Energy band structures of A$_3$B zigzag NRs for
the hopping ratios $\gamma^\prime/\gamma$ with (b) $1.0/1.5$, (c) $1.0$ and (d) $1.5$.
TES (red curves) appear in non-zero Zak's phase (yellow shaded region).
Blue shaded region indicates the ranges that Eqs.~(3) and (4) cannot be applied.}
 \label{fig:A3B_band_hop}
\end{figure*}

\subsection{Symmetry analysis}
Before showing the detailed results, we briefly look
at the symmetries of A$_3$B structure.
This structure has time-reversal and $C_{6v}$ point group
symmetries.
The two B atoms in unit cell play similar role of two nonequivalent
sublattices in graphene, since both of them are mutually transformed
under $C_6$ rotation. On the other hand,
the six A atoms in unit cell play similar role of benzene rings in
hexagonal 2D SSH model.~\cite{Liu2017B} Thus,
two of energy bands of A$_3$B resemble energy bands of graphene,
and the other six energy bands resemble energy bands of
hexagonal 2D SSH model discussed in Ref.~\onlinecite{Liu2017B}.

As there are both time-reversal and inversion symmetries, Berry
curvature in A$_3$B is guaranteed to vanish everywhere except energy-degenerate $K$ and
$K^\prime$ points in momentum space.
To apply Eqs.~(3) and (4), we lift these degeneracies by adding onsite
potentials $\delta_m\sim m\gamma/100$ ($m=1,2,3$) on A atoms to break $C_{6}$ point group
symmetry as shown in the inset of Fig.~2(a). After imposing such the
perturbation, degeneracies are lifted except for the central two
energy bands around $E=0$ at $K$ and $K^\prime$ points. Thus, when one of the central two energy bands is occupied,
Eqs.~(3) and (4) can be applied only for $-2\pi/3<k<2\pi/3$.

\subsection{Effect of variable hopping}
Figures~\ref{fig:2Dband} (a)-(c) show
energy band structures of A$_3$B sheet for different
ratios of $\gamma^\prime/\gamma$ with $V_A=V_B=0$.
The red and blue circles at $\Gamma$ and $M_1$ points indicate even
and odd parities of wavefunctions.
Yellow regions indicate that Zak's phase along $\mathbf{b}_1$
is $\pi$, resulting in charge polarization of $1/2$.

In Figs.~\ref{fig:2Dband} (a)-(c), $P^n_1(k=0)$ is calculated from the parities
of wavefunction for $n$-th energy band at $\Gamma$ and $M_1$ points by applying
Eq.~(3). Note that all three nonequivalent $M$ points have same
parity. Then we obtain Zak's phase at other $k$ points by applying Eq.~(4) except for the 
central two energy bands which have similar nature of graphene.



For the central two bands, they have
degenerate points $K$ and $K^\prime$ guaranteed by inversion symmetry. Thus, Equation~(4) is applied 
for a limited range $-2\pi/3<k<2\pi/3$ where gap-opening conditions are satisfied.
However, we can not apply Eq.~(4) for the path which go beyond the region $-2\pi/3<k<2\pi/3$ to relate
with the region $k<-2\pi/3$ or $k>2\pi/3$, because $F^n$ is not well defined at the degenerate 
points. In this situation, only Eq.~(2) is applicable for central two bands.~\cite{Delplace2011}
Thus, if one of these two central energy bands is
occupied, the value distribution of Zak's phase is dissociated into
two distinct regions as displayed by Fig.~1(b). Otherwise, Zak's phase
is uniform over whole BZ. 


Same procedure can be applied for $\mathbf{b}_2$ direction,
and $\mathbf{b}_1$ and $\mathbf{b}_2$ directions are equivalent owing
to $C_{6v}$ point group symmetry.


As we see from Fig.~\ref{fig:2Dband}, A$_3$B atomic layer always possesses finite
Zak's phase irrespective of the ratio between $\gamma^\prime$ and $\gamma$.
In case of $\gamma^\prime=\gamma$, Figure~\ref{fig:2Dband}(b) reproduces the energy band
structure of graphene, which possesses finite Zak's phase around zero
energy.~\cite{Delplace2011}
In the case of $\gamma^\prime<\gamma$ [Fig.~\ref{fig:2Dband}(a)], it
is similar to the case of $\gamma=\gamma^\prime$.
In the case of $\gamma<\gamma^\prime$ [Fig.~\ref{fig:2Dband}(c)],
besides graphene-like energy bands, upper and lower bands possess
finite Zak's phase due to band inversions. Thus,
emergence of TES is expected in A$_3$B system for any ratio between
$\gamma^\prime$ and $\gamma$.


To show TES induced by Zak's phase,
we study the energy band structures of A$_3$B NRs. Figure~\ref{fig:A3B_band_hop}(a) displays lattice
structure of A$_3$B NR with zigzag edges. For zigzag NR the corresponding Zak's
phase is along $\mathbf{b}_1$ or $\mathbf{b}_2$ direction. We assume
that all the edge atoms are terminated by
hydrogen atoms and no dangling bond exists. The width of NR is given
by number of zigzag chains $N_z$.

In Figs.~\ref{fig:A3B_band_hop}(b)-(d), we show the energy band structures of
NR for different hopping ratios $\gamma^\prime/\gamma$.
It can be clearly observed that TES appear in the subband gap regions
(indicated by yellow) where Zak's phases are $\pi$. In case of $\gamma^\prime/\gamma \le 1$, TES appear within the
central subband gap for zigzag NR.
For  $\gamma^\prime/\gamma > 1$,
band inversions occur in upper and lower energy regions away from $E=0$,
resulting in nonzero Zak's phases and consequent emergence of TES. It is noted that
the TES at E=0 only appear within the region $-2\pi/3<k<2\pi/3$ that
is same as graphene zigzag NR.~\cite{Delplace2011}
These zero energy edge states are nonbonding molecular
orbitals, whose analytic form can be derived in similar manners of
Refs.~\onlinecite{Fujita1996,Wakabayashi.jpsj.2010} as detailed in
Appendix~\ref{sec:appendix_estate}. Outside this range, Zak's phase
cannot be calculated by Eqs.~(3) and (4), which is shaded as blue.

\subsection{Effect of onsite potential}
\begin{figure}[t]
 \includegraphics[width=0.48\textwidth]{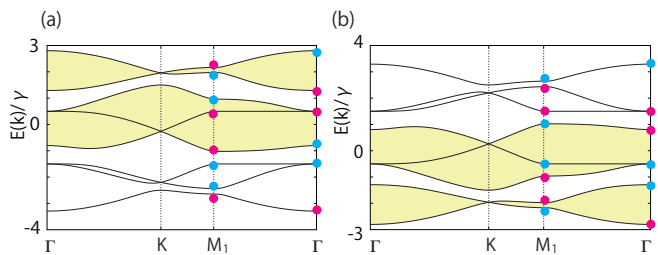}
\caption{Energy band structures of 2D A$_3$B biatomic sheet in presence of onsite potentials,
$V_A$ and $V_B$.
The parameter values are chosen as follows:
(a)$V_A=-0.5\gamma$, $V_B=0.5\gamma$,
(b)$V_A=0.5\gamma$, $V_B=-0.5\gamma$.
Here red (blue) circles at $\Gamma$ and $M_1$ points indicate the positive (negative) parity
of wavefunction. Yellow regions depict nonzero Zak's phase. Note that,
the hopping ratio $\gamma^{\prime}/\gamma$ is 1 for all plots.}
\label{fig:2Dband_onsite}
\end{figure}

In A$_3$B sheet, onsite potentials of A and B atoms are different
due to their distinct chemical elements. Here, we study effect of different onsite potentials on A and B atoms. The corresponding
energy band structures are displayed in Figs.~\ref{fig:2Dband_onsite}(a) and (b), where the
yellow regions indicate nonzero Zak's phase.

According to Fig.~\ref{fig:2Dband_onsite}, A$_3$B systems always possess nontrivial
energy bands in either the upper or lower energy region depending on
the values of $V_A$ and $V_B$. When $V_A<V_B$, the central and upper subband gaps have nonzero Zak's phase as shown in
Fig.~\ref{fig:2Dband_onsite}(a). For $V_A>V_B$, the central  and lower subband gaps have nonzero Zak's phase as shown in
Fig.~\ref{fig:2Dband_onsite}(b).
Thus, TES emerge in either upper or lower subband gaps when onsite potentials between
A and B atoms are different.

 Figures~\ref{fig:A3B_band_onsite}(a)-(c) show the energy band structures of
 A$_3$B zigzag NRs in presence of different onsite potentials between A and B atoms. TES appear in
 yellow shaded region where the Zak's phase is nonzero.
 In case of $V_A < V_B$, TES appear
 in the central and upper energy regions.
 When $V_A > V_B$, TES appear in the
 central and lower energy regions. In presence of different onsite
 potentials, note that TES emerge in
 the energy regions away from $E=0$ even when $\gamma^{\prime}=\gamma$.


\section{Density Functional Theory}
\label{sec:dft}
So far, on tight-binding calculations, we have demonstrated that A$_3$B sheet possesses nonzero Zak's
phase, which consequently induces TES. Here we investigate the
electronic structure of C$_3$N biatomic sheet as a realistic candidate
with nonzero Zak's phase on the basis of first-principles calculations using SIESTA.~\cite{siesta_method}

The conditions of first-principles calculations are summarized as follows.
Perdew-Burke-Ernzerhof (PBE) exchange and correlation
functional have been considered within generalized gradient
approximations with double zeta polarized (DZP) basis set. To avoid any
interactions within adjacent unit cells, we have created sufficiently
large vacuum regions in the non-periodic directions. The energy cut-off
for real space mesh size is 400 Ry energy.
The $k$-point sampling in BZ is
taken over $30\times30\times1$ of Monkhorst-Pack grid for the
relaxation of 2D C$_3$N sheet, and
$70\times1\times1$ for that of
C$_3$N zigzag NRs. The atomic positions are relaxed
until the force on each atom reaches 0.04
eV/\text{\AA}. For calculations of electronic states for optimized
NRs, we take the $k$-points sampling in BZ as
$300\times1\times1$ of Monkhorst-Pack grid.

\begin{figure}[t]
\begin{center}
\includegraphics[width=0.48\textwidth]{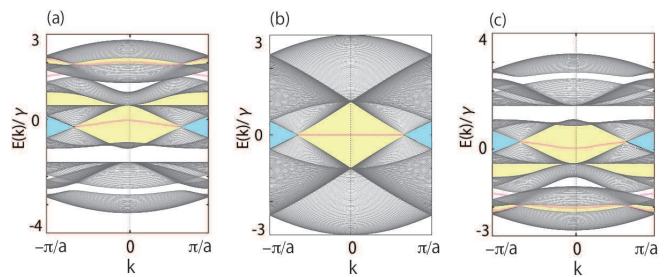}
\end{center}
\caption{Energy band structures of A$_3$B zigzag NR
for the following onsite potential values:
(a) $V_A=-0.5\gamma$ and $V_B=0.5\gamma$, (b) $V_A=0$ and $V_B=0$ and (c) $V_A=0.5\gamma$ and $V_B=-0.5\gamma$.
The yellow shaded region has nonzero
Zak's phase, where the TES (red curves) appear. Blue shaded region
indicates that Eqs.~(3) and (4) cannot be applied.
Note that, the hopping ratio $\gamma^{\prime}/\gamma$ is 1 for all plots.}
\label{fig:A3B_band_onsite}
\end{figure}

\begin{figure*}[ht]
\begin{center}
\includegraphics[width=0.9\textwidth]{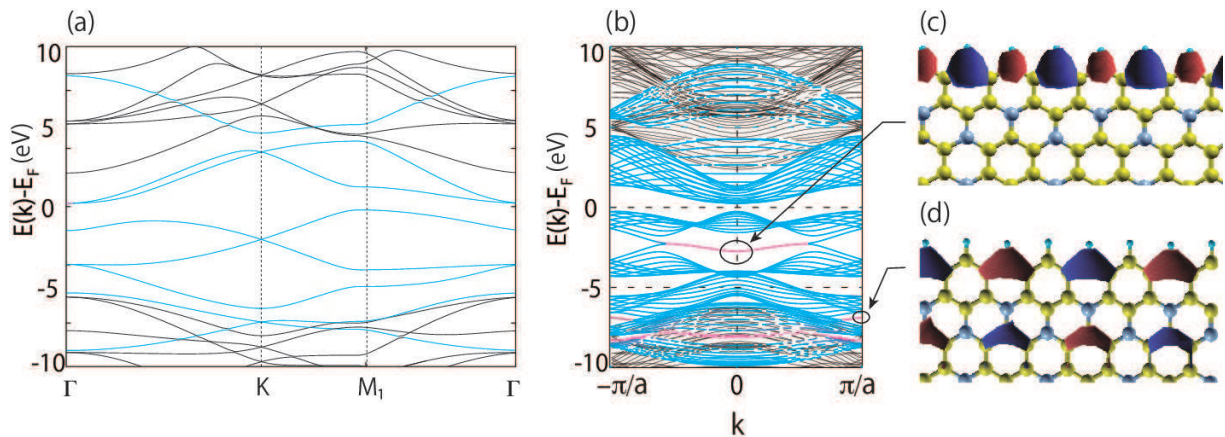}
\caption{(a) Energy band structure of C$_3$N sheet.
  Blue curves indicate $\pi$-electronic bands.
  (b) Energy band structure of C$_3$N zigzag NR.
  Blue and red curves indicate $\pi$-electronic bands. Especially,
  red curves indicate TES.
  Wavefunction of TES near the edge at energies of (c) $-2.5$eV and (d) $-7.0$eV.
   } \label{fig:c3n_dft}
\end{center}
\end{figure*}

Figure~\ref{fig:c3n_dft}(a) shows energy band structure of C$_3$N sheet,
where Fermi energy is zero.
Blue curves indicate the energy bands that originate from
$\pi$ electrons, which nicely match with the energy band structures
obtained by using tight-binding model shown in
Fig.~\ref{fig:2Dband_onsite}(b).
Since nitrogen has one excess electron than carbon, it should be noted that
the Fermi energy is upward shifted owing to the electron doping by
nitrogen substitution.
From Fig.~\ref{fig:2Dband_onsite}(b), the middle and low subband gaps possess finite Zak's phases. Thus,
TES induced by Zak's phase are expected to appear in C$_3$N.

Figure~\ref{fig:c3n_dft}(b) shows
energy band structure of C$_3$N zigzag NR with $N_z=20$.
Red and blue curves indicate the energy bands arising from $\pi$ electrons.
Especially, red curves indicate TES.
Edge states of partial flat bands appear near $-2.5$eV, $-7.0$eV and $-8.0$eV,
consistent with the tight-binding calculations [see Fig.~\ref{fig:A3B_band_onsite}(c)].
Wavefunction of the flat band near $-2.5$eV at $\Gamma$ point
is shown in Fig.~\ref{fig:c3n_dft}(c), which suggests strong localization of electrons near edges.
In addition, we also show wavefunction of the flat band
near $-7.0$eV at BZ boundary ($k=\frac{\pi}{a}$) in Fig.~\ref{fig:c3n_dft}(d), which
displays localized wavefunction at ribbon edges.

Thus, C$_3$N can be considered as one possible realistic
material that possess TES protected by nonzero Zak's phase.
Especially the TES near $-2.5$eV have similar electronic properties of edge states in 
zigzag graphene edges,~\cite{Fujita1996,Wakabayashi.jpsj.2010} they provide a perfectly electronic transport channel
which is robust to edge roughness and impurities as long as the
intervalley scattering are suppressed.~\cite{Wakabayashi2007,Wakabayashi2009A,Wakabayashi2009B}
Besides C$_3$N, we also investigate the electronic structure of
honeycomb BC$_3$ sheet on the basis of first-principles calculations.
The details are presented in Appendix~\ref{sec:mixture}.

\begin{figure}[ht]
\begin{center}
\includegraphics[width=0.45\textwidth]{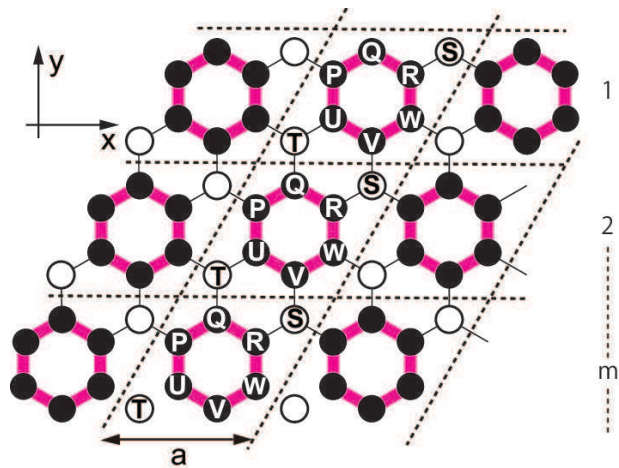}
\end{center}
\caption{Lattice structure of A$_3$B biatomic sheet.
Each unit cell contains eight sublattices denoted by {\it P,Q,R,S,T,U,V,W}. }
\label{fig:analytical}
\end{figure}

\begin{figure*}[ht]
 \includegraphics[width=0.9\textwidth]{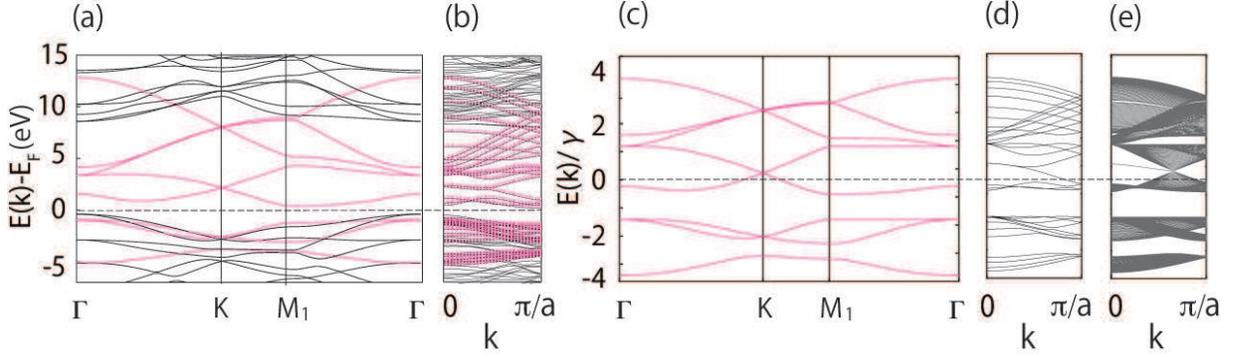}
\caption{
Energy band structures of (a) BC$_3$ sheet and (b) BC$_3$ zigzag NR
 ($N_z=8$) obtained by first-principles
 calculation. Red curves indicate $\pi$-electronic bands. The edges
 are terminated by hydrogen atoms. Energy band structures of (c) A$_3$B
 sheet and A$_3$B zigzag NR for (d)~$N_z=8$ and (e)~$N_z=100$, as
 obtained by using the tight-biding model. Here,
 $\gamma^\prime=\gamma/1.3$, $V_A=-0.5\gamma$,$V_B=0.5\gamma$.
}
\label{fig:mixture_bc3}
\end{figure*}

\section{summary}
\label{sec:summary}
 In summary, we have studied the electronic structures of A$_3$B
 biatomic sheet on the basis of tight-binding model. This system shows
 topological phase transition by tuning the electron hopping and onsite
 potentials. Instead of Berry curvature, this topological phase
 transition is characterized by non-zero Zak's phase, which induces
 TES.
 Based on our tight-binding analysis, we further
 propose realistic material candidates, {\it e.g.}, C$_3$N and BC$_3$. Within first-principles
 calculations we successfully demonstrate the emergence of TES in
 such materials.

\section{Acknowledgments}
S.D. and K.W. acknowledge the financial support from Hyogo Overseas
Research Network (HORN) and the financial support
for international collaboration of Kwansei Gakuin University.
F.L. is an overseas researcher under the Postdoctoral Fellowship of Japan
Society for the Promotion of Science (JSPS).
S.D. acknowledges Science and the Engineering Research Board (SERB),
Department of Science and Technology (DST), Government of India, for the
Early Career Research Award grant (ECR/2016/000283).
This work was supported by JSPS KAKENHI Grants No. JP25107005, No.
JP15K21722, No. JP15K13507, No. JP17F17326 and JP18H01154.

\appendix
\section{Edge states}
\label{sec:appendix_estate}
As we have shown in Figs.~\ref{fig:A3B_band_hop}(b)-(d), zigzag A$_3$B
NRs have the partial flat bands at $E=0$ for $|k|\le 2\pi/3$, where
electrons are localized near the edges, \textit{i.e.} edge states.
In this section, we derive the condition of $k$-region for which edge
states can exist in zigzag A$_3$B NRs by constructing an analytic
solution of edge state for semi-infinite A$_3$B sheet with a zigzag edge,
according to the manner presented in Refs.~\onlinecite{Fujita1996,Wakabayashi.jpsj.2010}.

In order to derive the wavefunctions of the edge states, we divide the graphene
lattice into eight sublattices, namely, {\it P,Q,R,T,U,V} and {\it W} as
shown in Fig.~\ref{fig:analytical}.
We define the wavefunction at {\it m}-th row as
\begin{eqnarray}
\Psi_{\it{m}}=(\psi_{\it{m,P}},\psi_{\it{m,Q}},\ldots,\psi_{\it{m,V}},\psi_{\it{m,W}}).
\end{eqnarray}
Here we have assumed the translational invariance along the ribbon direction.

Thus, the set of equation of motions for nearest-neighbor tight-binding model
can be written as
\begin{eqnarray}
  \epsilon\psi_{\it mP}&=&-\gamma^{\prime}\eta\psi_{\it mS}-\gamma\eta^{\ast}\psi_{\it mQ}-\gamma\eta^{\ast}\psi_{\it mU},\\
  \epsilon\psi_{\it mQ}&=&-\gamma^{\prime}\eta\psi_{\it m-1T}-\gamma\eta\psi_{\it mP}-\gamma\eta^{\ast}\psi_{\it mR},\\
  \epsilon\psi_{\it mR}&=&-\gamma\eta\psi_{\it mQ}-\gamma\eta^{\ast}\psi_{\it mW}-\gamma^{\prime}\eta^{\ast}\psi_{\it mS},\\
  \epsilon\psi_{\it mS}&=&-\gamma^{\prime}\eta\psi_{\it m-1V}-\gamma^{\prime}\eta\psi_{\it mR}-\gamma^{\prime}\eta^{\ast}\psi_{\it mS},\\
  \epsilon\psi_{\it mT}&=&-\gamma^{\prime}\eta\psi_{\it mW}-\gamma^{\prime}\eta^{\ast}\psi_{\it mU}-\gamma^{\prime}\eta^{\ast}\psi_{\it m+1Q},\\
  \epsilon\psi_{\it mU}&=&-\gamma\eta\psi_{\it mP}-\gamma^{\prime}\eta\psi_{\it mT}-\gamma\eta^{\ast}\psi_{\it mV},\\
  \epsilon\psi_{\it mV}&=&-\gamma\eta\psi_{\it mU}-\gamma\eta^{\ast}\psi_{\it mW}-\gamma^{\prime}\eta^{\ast}\psi_{\it m+1S},\\
  \epsilon\psi_{\it mW}&=&-\gamma\eta\psi_{\it mR}-\gamma\eta\psi_{\it mV}-\gamma^{\prime}\eta^{\ast}\psi_{\it mT}.
\end{eqnarray}
Here $\eta = \mathrm{e}^{ik/4} $ is the Bloch phase, and $\eta^{\ast}$
is its complex conjugate.
Here we have defined the lattice constant as unit of length.
Since we are interested in the wavefunctions of zero-mode
($\epsilon=0$),
we introduce the conditions:
\begin{eqnarray}
\epsilon=0,  \psi_{\it mP}=0, \psi_{\it mR}=0, \psi_{\it mT}=0, \psi_{\it mV}=0
\end{eqnarray}
We have numerically confirmed that the wavefunction at ({\it m,P}),
({\it m,R}), ({\it m,T}) and ({\it m,V}) sites for arbitrary {\it m}
is identically zero. Therefore, we can simplify the set of equation of motions:
\begin{eqnarray}
 0&=&-\gamma^{\prime}\eta\psi_{\it mS}-\gamma\eta^{\ast}\psi_{\it mQ}-\gamma\eta^{\ast}\psi_{\it mU},\\
 0&=&-\gamma\eta\psi_{\it mQ}-\gamma\eta^{\ast}\psi_{\it mW}-\gamma^{\prime}\eta^{\ast}\psi_{\it mS},\\
 0&=&-\gamma^{\prime}\eta\psi_{\it mW}-\gamma^{\prime}\eta^{\ast}\psi_{\it mU}-\gamma^{\prime}\eta^{\ast}\psi_{\it m+1Q},\\
 0&=&-\gamma\eta\psi_{\it mU}-\gamma\eta^{\ast}\psi_{\it mW}-\gamma^{\prime}\eta^{\ast}\psi_{\it m+1S}.
\end{eqnarray}
From these equations, the wavefunction at ($m,U$) and ($m,W$) sites can be
related to those at ($m,Q$) and ($m,S$) sites as following,
\begin{eqnarray}
  \psi_{\it mU}&=\eta\psi_{\it mQ}+\frac{\gamma^{\prime}}{\gamma}\eta^{\it \ast}\psi_{\it mS}\\
  \psi_{\it mW}&=\eta^{\it \ast}\psi_{\it mQ}+\frac{\gamma^{\prime}}{\gamma}\eta\psi_{\it mS}
\end{eqnarray}
Therefore we obtain the following recurrence equations for charge densities between adjacent cells,
\begin{eqnarray}
  |\psi_{\it m+1Q}|^{2}&=4|\psi_{mQ}|^{2}(1+\cos{k})^{2},\\
  |\psi_{\it m+1S}|^{2}&=4|\psi_{m+1S}|^{2}(1+\cos{k})^{2}.
\end{eqnarray}
Since the wavefunctions have to converge in the bulk region,
the prefactors of above equations have to satisfy the following condition,
\begin{equation}
  4(1+\cos{k/2})^{2}\leq 1.
\end{equation}
Immediately, we obtain the condition for wavenumber to satisfy $\epsilon=0$,
\begin{equation}
  -\frac{2\pi}{3}\leq k \leq\frac{2\pi}{3}.
\end{equation}
This is nothing more than the region of flat bands as shown
in Fig.~\ref{fig:analytical}.

\section{Mixture of hopping energy and onsite potential}
\label{sec:mixture}
In Sec.~\ref{sec:tbm}, we discussed the effect of hopping energy and onsite
potential separately.
Here, we take into account both hopping energy and onsite
potential simultaneously. The results well reproduce the energy band
structures of BC$_3$ sheet.~\cite{dutta_2012,dutta_2013}
Figure~\ref{fig:mixture_bc3} (a) and (b) show
the energy band structures of
BC$_3$ sheet and zigzag BC$_3$ NR ($N_z=8$), respectively, as obtained
within first-principles calculations. Here,
red curves indicate $\pi$ electron bands.
TES appear near $E=2.5$eV.
If we choose the parameters $\gamma^\prime=\gamma/1.3$,
$V_A=-0.5\gamma$,$V_B=0.5\gamma$ for tight binding model,
the energy band structure of BC$_3$ is well reproduced as shown in
Fig.~\ref{fig:mixture_bc3} (c).
Figures~\ref{fig:mixture_bc3} (d) and (e) are the 
energy band structures of A$_3$B zigzag NRs obtained by tight-binding model,
where TES clearly appear.

\bibliographystyle{apsrev4-1} 
\bibliography{references}

\begin{thebibliography}{44}%
\makeatletter
\providecommand \@ifxundefined [1]{%
 \@ifx{#1\undefined}
}%
\providecommand \@ifnum [1]{%
 \ifnum #1\expandafter \@firstoftwo
 \else \expandafter \@secondoftwo
 \fi
}%
\providecommand \@ifx [1]{%
 \ifx #1\expandafter \@firstoftwo
 \else \expandafter \@secondoftwo
 \fi
}%
\providecommand \natexlab [1]{#1}%
\providecommand \enquote  [1]{``#1''}%
\providecommand \bibnamefont  [1]{#1}%
\providecommand \bibfnamefont [1]{#1}%
\providecommand \citenamefont [1]{#1}%
\providecommand \href@noop [0]{\@secondoftwo}%
\providecommand \href [0]{\begingroup \@sanitize@url \@href}%
\providecommand \@href[1]{\@@startlink{#1}\@@href}%
\providecommand \@@href[1]{\endgroup#1\@@endlink}%
\providecommand \@sanitize@url [0]{\catcode `\\12\catcode `\$12\catcode
  `\&12\catcode `\#12\catcode `\^12\catcode `\_12\catcode `\%12\relax}%
\providecommand \@@startlink[1]{}%
\providecommand \@@endlink[0]{}%
\providecommand \url  [0]{\begingroup\@sanitize@url \@url }%
\providecommand \@url [1]{\endgroup\@href {#1}{\urlprefix }}%
\providecommand \urlprefix  [0]{URL }%
\providecommand \Eprint [0]{\href }%
\providecommand \doibase [0]{http://dx.doi.org/}%
\providecommand \selectlanguage [0]{\@gobble}%
\providecommand \bibinfo  [0]{\@secondoftwo}%
\providecommand \bibfield  [0]{\@secondoftwo}%
\providecommand \translation [1]{[#1]}%
\providecommand \BibitemOpen [0]{}%
\providecommand \bibitemStop [0]{}%
\providecommand \bibitemNoStop [0]{.\EOS\space}%
\providecommand \EOS [0]{\spacefactor3000\relax}%
\providecommand \BibitemShut  [1]{\csname bibitem#1\endcsname}%
\let\auto@bib@innerbib\@empty
\bibitem [{\citenamefont {Bansil}\ \emph {et~al.}(2016)\citenamefont {Bansil},
  \citenamefont {Lin},\ and\ \citenamefont {Das}}]{Bansil2016}%
  \BibitemOpen
  \bibfield  {author} {\bibinfo {author} {\bibfnamefont {A.}~\bibnamefont
  {Bansil}}, \bibinfo {author} {\bibfnamefont {H.}~\bibnamefont {Lin}}, \ and\
  \bibinfo {author} {\bibfnamefont {T.}~\bibnamefont {Das}},\ }\href {\doibase
  10.1103/RevModPhys.88.021004} {\bibfield  {journal} {\bibinfo  {journal}
  {Rev. Mod. Phys.}\ }\textbf {\bibinfo {volume} {88}},\ \bibinfo {pages}
  {021004} (\bibinfo {year} {2016})}\BibitemShut {NoStop}%
\bibitem [{\citenamefont {Kane}\ and\ \citenamefont {Mele}(2005)}]{Kane2005}%
  \BibitemOpen
  \bibfield  {author} {\bibinfo {author} {\bibfnamefont {C.~L.}\ \bibnamefont
  {Kane}}\ and\ \bibinfo {author} {\bibfnamefont {E.~J.}\ \bibnamefont
  {Mele}},\ }\href {\doibase 10.1103/PhysRevLett.95.226801} {\bibfield
  {journal} {\bibinfo  {journal} {Phys. Rev. Lett.}\ }\textbf {\bibinfo
  {volume} {95}},\ \bibinfo {pages} {226801} (\bibinfo {year}
  {2005})}\BibitemShut {NoStop}%
\bibitem [{\citenamefont {Bernevig}\ \emph {et~al.}(2006)\citenamefont
  {Bernevig}, \citenamefont {Hughes},\ and\ \citenamefont
  {Zhang}}]{Bernevig2006}%
  \BibitemOpen
  \bibfield  {author} {\bibinfo {author} {\bibfnamefont {B.~A.}\ \bibnamefont
  {Bernevig}}, \bibinfo {author} {\bibfnamefont {T.~L.}\ \bibnamefont
  {Hughes}}, \ and\ \bibinfo {author} {\bibfnamefont {S.-C.}\ \bibnamefont
  {Zhang}},\ }\href {\doibase 10.1126/science.1133734} {\bibfield  {journal}
  {\bibinfo  {journal} {Science}\ }\textbf {\bibinfo {volume} {314}},\ \bibinfo
  {pages} {1757} (\bibinfo {year} {2006})}\BibitemShut {NoStop}%
\bibitem [{\citenamefont {Hsieh}\ \emph {et~al.}(2008)\citenamefont {Hsieh},
  \citenamefont {Qian}, \citenamefont {Wray}, \citenamefont {Xia},
  \citenamefont {Hor}, \citenamefont {Cava},\ and\ \citenamefont
  {Hasan}}]{Hsieh2008}%
  \BibitemOpen
  \bibfield  {author} {\bibinfo {author} {\bibfnamefont {D.}~\bibnamefont
  {Hsieh}}, \bibinfo {author} {\bibfnamefont {D.}~\bibnamefont {Qian}},
  \bibinfo {author} {\bibfnamefont {L.}~\bibnamefont {Wray}}, \bibinfo {author}
  {\bibfnamefont {Y.}~\bibnamefont {Xia}}, \bibinfo {author} {\bibfnamefont
  {Y.~S.}\ \bibnamefont {Hor}}, \bibinfo {author} {\bibfnamefont {R.~J.}\
  \bibnamefont {Cava}}, \ and\ \bibinfo {author} {\bibfnamefont {M.~Z.}\
  \bibnamefont {Hasan}},\ }\href {\doibase 10.1038/nature06843} {\bibfield
  {journal} {\bibinfo  {journal} {Nature}\ }\textbf {\bibinfo {volume} {452}},\
  \bibinfo {pages} {970} (\bibinfo {year} {2008})}\BibitemShut {NoStop}%
\bibitem [{\citenamefont {Chen}\ \emph {et~al.}(2009)\citenamefont {Chen},
  \citenamefont {Analytis}, \citenamefont {Chu}, \citenamefont {Liu},
  \citenamefont {Mo}, \citenamefont {Qi}, \citenamefont {Zhang}, \citenamefont
  {Lu}, \citenamefont {Dai}, \citenamefont {Fang}, \citenamefont {Zhang},
  \citenamefont {Fisher}, \citenamefont {Hussain},\ and\ \citenamefont
  {Shen}}]{Chen2009}%
  \BibitemOpen
  \bibfield  {author} {\bibinfo {author} {\bibfnamefont {Y.~L.}\ \bibnamefont
  {Chen}}, \bibinfo {author} {\bibfnamefont {J.~G.}\ \bibnamefont {Analytis}},
  \bibinfo {author} {\bibfnamefont {J.-H.}\ \bibnamefont {Chu}}, \bibinfo
  {author} {\bibfnamefont {Z.~K.}\ \bibnamefont {Liu}}, \bibinfo {author}
  {\bibfnamefont {S.-K.}\ \bibnamefont {Mo}}, \bibinfo {author} {\bibfnamefont
  {X.~L.}\ \bibnamefont {Qi}}, \bibinfo {author} {\bibfnamefont {H.~J.}\
  \bibnamefont {Zhang}}, \bibinfo {author} {\bibfnamefont {D.~H.}\ \bibnamefont
  {Lu}}, \bibinfo {author} {\bibfnamefont {X.}~\bibnamefont {Dai}}, \bibinfo
  {author} {\bibfnamefont {Z.}~\bibnamefont {Fang}}, \bibinfo {author}
  {\bibfnamefont {S.~C.}\ \bibnamefont {Zhang}}, \bibinfo {author}
  {\bibfnamefont {I.~R.}\ \bibnamefont {Fisher}}, \bibinfo {author}
  {\bibfnamefont {Z.}~\bibnamefont {Hussain}}, \ and\ \bibinfo {author}
  {\bibfnamefont {Z.-X.}\ \bibnamefont {Shen}},\ }\href {\doibase
  10.1126/science.1173034} {\bibfield  {journal} {\bibinfo  {journal}
  {Science}\ }\textbf {\bibinfo {volume} {325}},\ \bibinfo {pages} {178}
  (\bibinfo {year} {2009})}\BibitemShut {NoStop}%
\bibitem [{\citenamefont {Chang}\ \emph {et~al.}(2013)\citenamefont {Chang},
  \citenamefont {Zhang}, \citenamefont {Feng}, \citenamefont {Shen},
  \citenamefont {Zhang}, \citenamefont {Guo}, \citenamefont {Li}, \citenamefont
  {Ou}, \citenamefont {Wei}, \citenamefont {Wang}, \citenamefont {Ji},
  \citenamefont {Feng}, \citenamefont {Ji}, \citenamefont {Chen}, \citenamefont
  {Jia}, \citenamefont {Dai}, \citenamefont {Fang}, \citenamefont {Zhang},
  \citenamefont {He}, \citenamefont {Wang}, \citenamefont {Lu}, \citenamefont
  {Ma},\ and\ \citenamefont {Xue}}]{Chang2013}%
  \BibitemOpen
  \bibfield  {author} {\bibinfo {author} {\bibfnamefont {C.-Z.}\ \bibnamefont
  {Chang}}, \bibinfo {author} {\bibfnamefont {J.}~\bibnamefont {Zhang}},
  \bibinfo {author} {\bibfnamefont {X.}~\bibnamefont {Feng}}, \bibinfo {author}
  {\bibfnamefont {J.}~\bibnamefont {Shen}}, \bibinfo {author} {\bibfnamefont
  {Z.}~\bibnamefont {Zhang}}, \bibinfo {author} {\bibfnamefont
  {M.}~\bibnamefont {Guo}}, \bibinfo {author} {\bibfnamefont {K.}~\bibnamefont
  {Li}}, \bibinfo {author} {\bibfnamefont {Y.}~\bibnamefont {Ou}}, \bibinfo
  {author} {\bibfnamefont {P.}~\bibnamefont {Wei}}, \bibinfo {author}
  {\bibfnamefont {L.-L.}\ \bibnamefont {Wang}}, \bibinfo {author}
  {\bibfnamefont {Z.-Q.}\ \bibnamefont {Ji}}, \bibinfo {author} {\bibfnamefont
  {Y.}~\bibnamefont {Feng}}, \bibinfo {author} {\bibfnamefont {S.}~\bibnamefont
  {Ji}}, \bibinfo {author} {\bibfnamefont {X.}~\bibnamefont {Chen}}, \bibinfo
  {author} {\bibfnamefont {J.}~\bibnamefont {Jia}}, \bibinfo {author}
  {\bibfnamefont {X.}~\bibnamefont {Dai}}, \bibinfo {author} {\bibfnamefont
  {Z.}~\bibnamefont {Fang}}, \bibinfo {author} {\bibfnamefont {S.-C.}\
  \bibnamefont {Zhang}}, \bibinfo {author} {\bibfnamefont {K.}~\bibnamefont
  {He}}, \bibinfo {author} {\bibfnamefont {Y.}~\bibnamefont {Wang}}, \bibinfo
  {author} {\bibfnamefont {L.}~\bibnamefont {Lu}}, \bibinfo {author}
  {\bibfnamefont {X.-C.}\ \bibnamefont {Ma}}, \ and\ \bibinfo {author}
  {\bibfnamefont {Q.-K.}\ \bibnamefont {Xue}},\ }\href {\doibase
  10.1126/science.1234414} {\bibfield  {journal} {\bibinfo  {journal}
  {Science}\ }\textbf {\bibinfo {volume} {340}},\ \bibinfo {pages} {167}
  (\bibinfo {year} {2013})}\BibitemShut {NoStop}%
\bibitem [{\citenamefont {Ando}(2013)}]{Ando2013}%
  \BibitemOpen
  \bibfield  {author} {\bibinfo {author} {\bibfnamefont {Y.}~\bibnamefont
  {Ando}},\ }\href {\doibase 10.7566/JPSJ.82.102001} {\bibfield  {journal}
  {\bibinfo  {journal} {J. Phys. Soc. Jpn.}\ }\textbf {\bibinfo {volume}
  {82}},\ \bibinfo {pages} {102001} (\bibinfo {year} {2013})}\BibitemShut
  {NoStop}%
\bibitem [{\citenamefont {Sato}\ and\ \citenamefont
  {Fujimoto}(2016)}]{Sato2016}%
  \BibitemOpen
  \bibfield  {author} {\bibinfo {author} {\bibfnamefont {M.}~\bibnamefont
  {Sato}}\ and\ \bibinfo {author} {\bibfnamefont {S.}~\bibnamefont
  {Fujimoto}},\ }\href {\doibase 10.7566/JPSJ.85.072001} {\bibfield  {journal}
  {\bibinfo  {journal} {J. Phys. Soc. Jpn.}\ }\textbf {\bibinfo {volume}
  {85}},\ \bibinfo {pages} {072001} (\bibinfo {year} {2016})}\BibitemShut
  {NoStop}%
\bibitem [{\citenamefont {Fu}(2011)}]{Fu2011}%
  \BibitemOpen
  \bibfield  {author} {\bibinfo {author} {\bibfnamefont {L.}~\bibnamefont
  {Fu}},\ }\href {\doibase 10.1103/PhysRevLett.106.106802} {\bibfield
  {journal} {\bibinfo  {journal} {Phys. Rev. Lett.}\ }\textbf {\bibinfo
  {volume} {106}},\ \bibinfo {pages} {106802} (\bibinfo {year}
  {2011})}\BibitemShut {NoStop}%
\bibitem [{\citenamefont {Tanaka}\ \emph {et~al.}(2012)\citenamefont {Tanaka},
  \citenamefont {Ren}, \citenamefont {Sato}, \citenamefont {Nakayama},
  \citenamefont {Souma}, \citenamefont {Takahashi}, \citenamefont {Segawa},\
  and\ \citenamefont {Ando}}]{Tanaka2012}%
  \BibitemOpen
  \bibfield  {author} {\bibinfo {author} {\bibfnamefont {Y.}~\bibnamefont
  {Tanaka}}, \bibinfo {author} {\bibfnamefont {Z.}~\bibnamefont {Ren}},
  \bibinfo {author} {\bibfnamefont {T.}~\bibnamefont {Sato}}, \bibinfo {author}
  {\bibfnamefont {K.}~\bibnamefont {Nakayama}}, \bibinfo {author}
  {\bibfnamefont {S.}~\bibnamefont {Souma}}, \bibinfo {author} {\bibfnamefont
  {T.}~\bibnamefont {Takahashi}}, \bibinfo {author} {\bibfnamefont
  {K.}~\bibnamefont {Segawa}}, \ and\ \bibinfo {author} {\bibfnamefont
  {Y.}~\bibnamefont {Ando}},\ }\href {http://dx.doi.org/10.1038/nphys2442}
  {\bibfield  {journal} {\bibinfo  {journal} {Nat. Phys.}\ }\textbf {\bibinfo
  {volume} {8}},\ \bibinfo {pages} {800} (\bibinfo {year} {2012})}\BibitemShut
  {NoStop}%
\bibitem [{\citenamefont {Dziawa}\ \emph {et~al.}(2012)\citenamefont {Dziawa},
  \citenamefont {Kowalski}, \citenamefont {Dybko}, \citenamefont {Buczko},
  \citenamefont {Szczerbakow}, \citenamefont {Szot}, \citenamefont
  {{\L}usakowska}, \citenamefont {Balasubramanian}, \citenamefont {Wojek},
  \citenamefont {Berntsen}, \citenamefont {Tjernberg},\ and\ \citenamefont
  {Story}}]{Dziawa2012a}%
  \BibitemOpen
  \bibfield  {author} {\bibinfo {author} {\bibfnamefont {P.}~\bibnamefont
  {Dziawa}}, \bibinfo {author} {\bibfnamefont {B.~J.}\ \bibnamefont
  {Kowalski}}, \bibinfo {author} {\bibfnamefont {K.}~\bibnamefont {Dybko}},
  \bibinfo {author} {\bibfnamefont {R.}~\bibnamefont {Buczko}}, \bibinfo
  {author} {\bibfnamefont {A.}~\bibnamefont {Szczerbakow}}, \bibinfo {author}
  {\bibfnamefont {M.}~\bibnamefont {Szot}}, \bibinfo {author} {\bibfnamefont
  {E.}~\bibnamefont {{\L}usakowska}}, \bibinfo {author} {\bibfnamefont
  {T.}~\bibnamefont {Balasubramanian}}, \bibinfo {author} {\bibfnamefont
  {B.~M.}\ \bibnamefont {Wojek}}, \bibinfo {author} {\bibfnamefont {M.~H.}\
  \bibnamefont {Berntsen}}, \bibinfo {author} {\bibfnamefont {O.}~\bibnamefont
  {Tjernberg}}, \ and\ \bibinfo {author} {\bibfnamefont {T.}~\bibnamefont
  {Story}},\ }\href {http://dx.doi.org/10.1038/nmat3449} {\bibfield  {journal}
  {\bibinfo  {journal} {Nat. Mater.}\ }\textbf {\bibinfo {volume} {11}},\
  \bibinfo {pages} {1023} (\bibinfo {year} {2012})}\BibitemShut {NoStop}%
\bibitem [{\citenamefont {Wan}\ \emph {et~al.}(2011)\citenamefont {Wan},
  \citenamefont {Turner}, \citenamefont {Vishwanath},\ and\ \citenamefont
  {Savrasov}}]{Wan2011}%
  \BibitemOpen
  \bibfield  {author} {\bibinfo {author} {\bibfnamefont {X.}~\bibnamefont
  {Wan}}, \bibinfo {author} {\bibfnamefont {A.~M.}\ \bibnamefont {Turner}},
  \bibinfo {author} {\bibfnamefont {A.}~\bibnamefont {Vishwanath}}, \ and\
  \bibinfo {author} {\bibfnamefont {S.~Y.}\ \bibnamefont {Savrasov}},\ }\href
  {\doibase 10.1103/PhysRevB.83.205101} {\bibfield  {journal} {\bibinfo
  {journal} {Phys. Rev. B}\ }\textbf {\bibinfo {volume} {83}},\ \bibinfo
  {pages} {205101} (\bibinfo {year} {2011})}\BibitemShut {NoStop}%
\bibitem [{\citenamefont {Borisenko}\ \emph {et~al.}(2014)\citenamefont
  {Borisenko}, \citenamefont {Gibson}, \citenamefont {Evtushinsky},
  \citenamefont {Zabolotnyy}, \citenamefont {B\"uchner},\ and\ \citenamefont
  {Cava}}]{Borisenko2014}%
  \BibitemOpen
  \bibfield  {author} {\bibinfo {author} {\bibfnamefont {S.}~\bibnamefont
  {Borisenko}}, \bibinfo {author} {\bibfnamefont {Q.}~\bibnamefont {Gibson}},
  \bibinfo {author} {\bibfnamefont {D.}~\bibnamefont {Evtushinsky}}, \bibinfo
  {author} {\bibfnamefont {V.}~\bibnamefont {Zabolotnyy}}, \bibinfo {author}
  {\bibfnamefont {B.}~\bibnamefont {B\"uchner}}, \ and\ \bibinfo {author}
  {\bibfnamefont {R.~J.}\ \bibnamefont {Cava}},\ }\href {\doibase
  10.1103/PhysRevLett.113.027603} {\bibfield  {journal} {\bibinfo  {journal}
  {Phys. Rev. Lett.}\ }\textbf {\bibinfo {volume} {113}},\ \bibinfo {pages}
  {027603} (\bibinfo {year} {2014})}\BibitemShut {NoStop}%
\bibitem [{\citenamefont {Lu}\ \emph {et~al.}(2015)\citenamefont {Lu},
  \citenamefont {Wang}, \citenamefont {Ye}, \citenamefont {Ran}, \citenamefont
  {Fu}, \citenamefont {Joannopoulos},\ and\ \citenamefont
  {Solja{\v{c}}i{\'{c}}}}]{Lu2015}%
  \BibitemOpen
  \bibfield  {author} {\bibinfo {author} {\bibfnamefont {L.}~\bibnamefont
  {Lu}}, \bibinfo {author} {\bibfnamefont {Z.}~\bibnamefont {Wang}}, \bibinfo
  {author} {\bibfnamefont {D.}~\bibnamefont {Ye}}, \bibinfo {author}
  {\bibfnamefont {L.}~\bibnamefont {Ran}}, \bibinfo {author} {\bibfnamefont
  {L.}~\bibnamefont {Fu}}, \bibinfo {author} {\bibfnamefont {J.~D.}\
  \bibnamefont {Joannopoulos}}, \ and\ \bibinfo {author} {\bibfnamefont
  {M.}~\bibnamefont {Solja{\v{c}}i{\'{c}}}},\ }\href {\doibase
  10.1126/science.aaa9273} {\bibfield  {journal} {\bibinfo  {journal}
  {Science}\ }\textbf {\bibinfo {volume} {349}},\ \bibinfo {pages} {622}
  (\bibinfo {year} {2015})}\BibitemShut {NoStop}%
\bibitem [{\citenamefont {Checkelsky}\ \emph {et~al.}(2012)\citenamefont
  {Checkelsky}, \citenamefont {Ye}, \citenamefont {Onose}, \citenamefont
  {Iwasa},\ and\ \citenamefont {Tokura}}]{Checkelsky2012}%
  \BibitemOpen
  \bibfield  {author} {\bibinfo {author} {\bibfnamefont {J.~G.}\ \bibnamefont
  {Checkelsky}}, \bibinfo {author} {\bibfnamefont {J.}~\bibnamefont {Ye}},
  \bibinfo {author} {\bibfnamefont {Y.}~\bibnamefont {Onose}}, \bibinfo
  {author} {\bibfnamefont {Y.}~\bibnamefont {Iwasa}}, \ and\ \bibinfo {author}
  {\bibfnamefont {Y.}~\bibnamefont {Tokura}},\ }\href
  {http://dx.doi.org/10.1038/nphys2388} {\bibfield  {journal} {\bibinfo
  {journal} {Nat. Phys.}\ }\textbf {\bibinfo {volume} {8}},\ \bibinfo {pages}
  {729} (\bibinfo {year} {2012})}\BibitemShut {NoStop}%
\bibitem [{\citenamefont {Xiao}\ \emph {et~al.}(2010)\citenamefont {Xiao},
  \citenamefont {Chang},\ and\ \citenamefont {Niu}}]{Xiao2010}%
  \BibitemOpen
  \bibfield  {author} {\bibinfo {author} {\bibfnamefont {D.}~\bibnamefont
  {Xiao}}, \bibinfo {author} {\bibfnamefont {M.-C.}\ \bibnamefont {Chang}}, \
  and\ \bibinfo {author} {\bibfnamefont {Q.}~\bibnamefont {Niu}},\ }\href
  {\doibase 10.1103/RevModPhys.82.1959} {\bibfield  {journal} {\bibinfo
  {journal} {Rev. Mod. Phys.}\ }\textbf {\bibinfo {volume} {82}},\ \bibinfo
  {pages} {1959} (\bibinfo {year} {2010})}\BibitemShut {NoStop}%
\bibitem [{\citenamefont {Sheng}\ \emph {et~al.}(2006)\citenamefont {Sheng},
  \citenamefont {Weng}, \citenamefont {Sheng},\ and\ \citenamefont
  {Haldane}}]{Haldane2006}%
  \BibitemOpen
  \bibfield  {author} {\bibinfo {author} {\bibfnamefont {D.~N.}\ \bibnamefont
  {Sheng}}, \bibinfo {author} {\bibfnamefont {Z.~Y.}\ \bibnamefont {Weng}},
  \bibinfo {author} {\bibfnamefont {L.}~\bibnamefont {Sheng}}, \ and\ \bibinfo
  {author} {\bibfnamefont {F.~D.~M.}\ \bibnamefont {Haldane}},\ }\href
  {\doibase 10.1103/PhysRevLett.97.036808} {\bibfield  {journal} {\bibinfo
  {journal} {Phys. Rev. Lett.}\ }\textbf {\bibinfo {volume} {97}},\ \bibinfo
  {pages} {036808} (\bibinfo {year} {2006})}\BibitemShut {NoStop}%
\bibitem [{\citenamefont {Liu}\ and\ \citenamefont
  {Wakabayashi}(2017)}]{Liu2017}%
  \BibitemOpen
  \bibfield  {author} {\bibinfo {author} {\bibfnamefont {F.}~\bibnamefont
  {Liu}}\ and\ \bibinfo {author} {\bibfnamefont {K.}~\bibnamefont
  {Wakabayashi}},\ }\href {\doibase 10.1103/PhysRevLett.118.076803} {\bibfield
  {journal} {\bibinfo  {journal} {Phys. Rev. Lett.}\ }\textbf {\bibinfo
  {volume} {118}},\ \bibinfo {pages} {076803} (\bibinfo {year}
  {2017})}\BibitemShut {NoStop}%
\bibitem [{\citenamefont {Liu}\ \emph {et~al.}(2017)\citenamefont {Liu},
  \citenamefont {Yamamoto},\ and\ \citenamefont {Wakabayashi}}]{Liu2017B}%
  \BibitemOpen
  \bibfield  {author} {\bibinfo {author} {\bibfnamefont {F.}~\bibnamefont
  {Liu}}, \bibinfo {author} {\bibfnamefont {M.}~\bibnamefont {Yamamoto}}, \
  and\ \bibinfo {author} {\bibfnamefont {K.}~\bibnamefont {Wakabayashi}},\
  }\href {\doibase 10.7566/JPSJ.86.123707} {\bibfield  {journal} {\bibinfo
  {journal} {J. Phys. Soc. Jpn.}\ }\textbf {\bibinfo {volume} {86}},\ \bibinfo
  {pages} {123707} (\bibinfo {year} {2017})}\BibitemShut {NoStop}%
\bibitem [{\citenamefont {Liu}\ \emph {et~al.}(2018)\citenamefont {Liu},
  \citenamefont {Deng},\ and\ \citenamefont {Wakabayashi}}]{Liu2018}%
  \BibitemOpen
  \bibfield  {author} {\bibinfo {author} {\bibfnamefont {F.}~\bibnamefont
  {Liu}}, \bibinfo {author} {\bibfnamefont {H.-Y.}\ \bibnamefont {Deng}}, \
  and\ \bibinfo {author} {\bibfnamefont {K.}~\bibnamefont {Wakabayashi}},\
  }\href {\doibase 10.1103/PhysRevB.97.035442} {\bibfield  {journal} {\bibinfo
  {journal} {Phys. Rev. B}\ }\textbf {\bibinfo {volume} {97}},\ \bibinfo
  {pages} {035442} (\bibinfo {year} {2018})}\BibitemShut {NoStop}%
\bibitem [{\citenamefont {Zak}(1989)}]{Zak1989}%
  \BibitemOpen
  \bibfield  {author} {\bibinfo {author} {\bibfnamefont {J.}~\bibnamefont
  {Zak}},\ }\href {\doibase 10.1103/PhysRevLett.62.2747} {\bibfield  {journal}
  {\bibinfo  {journal} {Phys. Rev. Lett.}\ }\textbf {\bibinfo {volume} {62}},\
  \bibinfo {pages} {2747} (\bibinfo {year} {1989})}\BibitemShut {NoStop}%
\bibitem [{\citenamefont {Delplace}\ \emph {et~al.}(2011)\citenamefont
  {Delplace}, \citenamefont {Ullmo},\ and\ \citenamefont
  {Montambaux}}]{Delplace2011}%
  \BibitemOpen
  \bibfield  {author} {\bibinfo {author} {\bibfnamefont {P.}~\bibnamefont
  {Delplace}}, \bibinfo {author} {\bibfnamefont {D.}~\bibnamefont {Ullmo}}, \
  and\ \bibinfo {author} {\bibfnamefont {G.}~\bibnamefont {Montambaux}},\
  }\href {\doibase 10.1103/PhysRevB.84.195452} {\bibfield  {journal} {\bibinfo
  {journal} {Phys. Rev. B}\ }\textbf {\bibinfo {volume} {84}},\ \bibinfo
  {pages} {195452} (\bibinfo {year} {2011})}\BibitemShut {NoStop}%
\bibitem [{\citenamefont {King-Smith}\ and\ \citenamefont
  {Vanderbilt}(1993)}]{King1993}%
  \BibitemOpen
  \bibfield  {author} {\bibinfo {author} {\bibfnamefont {R.~D.}\ \bibnamefont
  {King-Smith}}\ and\ \bibinfo {author} {\bibfnamefont {D.}~\bibnamefont
  {Vanderbilt}},\ }\href {\doibase 10.1103/PhysRevB.47.1651} {\bibfield
  {journal} {\bibinfo  {journal} {Phys. Rev. B}\ }\textbf {\bibinfo {volume}
  {47}},\ \bibinfo {pages} {1651} (\bibinfo {year} {1993})}\BibitemShut
  {NoStop}%
\bibitem [{\citenamefont {Resta}(1994)}]{Resta1994}%
  \BibitemOpen
  \bibfield  {author} {\bibinfo {author} {\bibfnamefont {R.}~\bibnamefont
  {Resta}},\ }\href {\doibase 10.1103/RevModPhys.66.899} {\bibfield  {journal}
  {\bibinfo  {journal} {Rev. Mod. Phys.}\ }\textbf {\bibinfo {volume} {66}},\
  \bibinfo {pages} {899} (\bibinfo {year} {1994})}\BibitemShut {NoStop}%
\bibitem [{\citenamefont {Zhou}\ \emph {et~al.}(2015)\citenamefont {Zhou},
  \citenamefont {Rabe},\ and\ \citenamefont {Vanderbilt}}]{Zhou2015}%
  \BibitemOpen
  \bibfield  {author} {\bibinfo {author} {\bibfnamefont {Y.}~\bibnamefont
  {Zhou}}, \bibinfo {author} {\bibfnamefont {K.~M.}\ \bibnamefont {Rabe}}, \
  and\ \bibinfo {author} {\bibfnamefont {D.}~\bibnamefont {Vanderbilt}},\
  }\href {\doibase 10.1103/PhysRevB.92.041102} {\bibfield  {journal} {\bibinfo
  {journal} {Phys. Rev. B}\ }\textbf {\bibinfo {volume} {92}},\ \bibinfo
  {pages} {041102} (\bibinfo {year} {2015})}\BibitemShut {NoStop}%
\bibitem [{\citenamefont {Huang}\ \emph {et~al.}(2018)\citenamefont {Huang},
  \citenamefont {Jin}, \citenamefont {Zhang},\ and\ \citenamefont
  {Liu}}]{Huang2018}%
  \BibitemOpen
  \bibfield  {author} {\bibinfo {author} {\bibfnamefont {H.}~\bibnamefont
  {Huang}}, \bibinfo {author} {\bibfnamefont {K.-H.}\ \bibnamefont {Jin}},
  \bibinfo {author} {\bibfnamefont {S.}~\bibnamefont {Zhang}}, \ and\ \bibinfo
  {author} {\bibfnamefont {F.}~\bibnamefont {Liu}},\ }\href@noop {} {\bibfield
  {journal} {\bibinfo  {journal} {Nano Lett.}\ }\textbf {\bibinfo {volume}
  {18}},\ \bibinfo {pages} {1972} (\bibinfo {year} {2018})}\BibitemShut
  {NoStop}%
\bibitem [{\citenamefont {Hirayama}\ \emph {et~al.}(2018)\citenamefont
  {Hirayama}, \citenamefont {Matsuishi}, \citenamefont {Hosono},\ and\
  \citenamefont {Murakami}}]{Hirayama2018}%
  \BibitemOpen
  \bibfield  {author} {\bibinfo {author} {\bibfnamefont {M.}~\bibnamefont
  {Hirayama}}, \bibinfo {author} {\bibfnamefont {S.}~\bibnamefont {Matsuishi}},
  \bibinfo {author} {\bibfnamefont {H.}~\bibnamefont {Hosono}}, \ and\ \bibinfo
  {author} {\bibfnamefont {S.}~\bibnamefont {Murakami}},\ }\href {\doibase
  10.1103/PhysRevX.8.031067} {\bibfield  {journal} {\bibinfo  {journal} {Phys.
  Rev. X}\ }\textbf {\bibinfo {volume} {8}},\ \bibinfo {pages} {031067}
  (\bibinfo {year} {2018})}\BibitemShut {NoStop}%
\bibitem [{\citenamefont {Heeger}\ \emph {et~al.}(1988)\citenamefont {Heeger},
  \citenamefont {Kivelson}, \citenamefont {Schrieffer},\ and\ \citenamefont
  {Su}}]{Heeger1988}%
  \BibitemOpen
  \bibfield  {author} {\bibinfo {author} {\bibfnamefont {A.~J.}\ \bibnamefont
  {Heeger}}, \bibinfo {author} {\bibfnamefont {S.}~\bibnamefont {Kivelson}},
  \bibinfo {author} {\bibfnamefont {J.~R.}\ \bibnamefont {Schrieffer}}, \ and\
  \bibinfo {author} {\bibfnamefont {W.~P.}\ \bibnamefont {Su}},\ }\href
  {\doibase 10.1103/RevModPhys.60.781} {\bibfield  {journal} {\bibinfo
  {journal} {Rev. Mod. Phys.}\ }\textbf {\bibinfo {volume} {60}},\ \bibinfo
  {pages} {781} (\bibinfo {year} {1988})}\BibitemShut {NoStop}%
\bibitem [{\citenamefont {Yanagisawa}\ \emph {et~al.}(2004)\citenamefont
  {Yanagisawa}, \citenamefont {Tanaka}, \citenamefont {Ishida}, \citenamefont
  {Matsue}, \citenamefont {Rokuta}, \citenamefont {Otani},\ and\ \citenamefont
  {Oshima}}]{Yanagisawa2004}%
  \BibitemOpen
  \bibfield  {author} {\bibinfo {author} {\bibfnamefont {H.}~\bibnamefont
  {Yanagisawa}}, \bibinfo {author} {\bibfnamefont {T.}~\bibnamefont {Tanaka}},
  \bibinfo {author} {\bibfnamefont {Y.}~\bibnamefont {Ishida}}, \bibinfo
  {author} {\bibfnamefont {M.}~\bibnamefont {Matsue}}, \bibinfo {author}
  {\bibfnamefont {E.}~\bibnamefont {Rokuta}}, \bibinfo {author} {\bibfnamefont
  {S.}~\bibnamefont {Otani}}, \ and\ \bibinfo {author} {\bibfnamefont
  {C.}~\bibnamefont {Oshima}},\ }\href {\doibase 10.1103/PhysRevLett.93.177003}
  {\bibfield  {journal} {\bibinfo  {journal} {Phys. Rev. Lett.}\ }\textbf
  {\bibinfo {volume} {93}},\ \bibinfo {pages} {177003} (\bibinfo {year}
  {2004})}\BibitemShut {NoStop}%
\bibitem [{\citenamefont {Yanagisawa}\ \emph {et~al.}(2006)\citenamefont
  {Yanagisawa}, \citenamefont {Ishida}, \citenamefont {Tanaka}, \citenamefont
  {Ueno}, \citenamefont {Otani},\ and\ \citenamefont {Oshima}}]{Tanaka2006}%
  \BibitemOpen
  \bibfield  {author} {\bibinfo {author} {\bibfnamefont {H.}~\bibnamefont
  {Yanagisawa}}, \bibinfo {author} {\bibfnamefont {Y.}~\bibnamefont {Ishida}},
  \bibinfo {author} {\bibfnamefont {T.}~\bibnamefont {Tanaka}}, \bibinfo
  {author} {\bibfnamefont {A.}~\bibnamefont {Ueno}}, \bibinfo {author}
  {\bibfnamefont {S.}~\bibnamefont {Otani}}, \ and\ \bibinfo {author}
  {\bibfnamefont {C.}~\bibnamefont {Oshima}},\ }\href {\doibase
  10.1016/j.susc.2006.01.124} {\bibfield  {journal} {\bibinfo  {journal} {Surf.
  Sci.}\ }\textbf {\bibinfo {volume} {600}},\ \bibinfo {pages} {4072} (\bibinfo
  {year} {2006})}\BibitemShut {NoStop}%
\bibitem [{\citenamefont {Mahmood}\ \emph {et~al.}(2016)\citenamefont
  {Mahmood}, \citenamefont {Lee}, \citenamefont {Jung}, \citenamefont {Shin},
  \citenamefont {Choi}, \citenamefont {Seo}, \citenamefont {Jung},
  \citenamefont {Kim}, \citenamefont {Li}, \citenamefont {Lah}, \citenamefont
  {Park}, \citenamefont {Shin}, \citenamefont {Oh},\ and\ \citenamefont
  {Baek}}]{Mahmood2016}%
  \BibitemOpen
  \bibfield  {author} {\bibinfo {author} {\bibfnamefont {J.}~\bibnamefont
  {Mahmood}}, \bibinfo {author} {\bibfnamefont {E.~K.}\ \bibnamefont {Lee}},
  \bibinfo {author} {\bibfnamefont {M.}~\bibnamefont {Jung}}, \bibinfo {author}
  {\bibfnamefont {D.}~\bibnamefont {Shin}}, \bibinfo {author} {\bibfnamefont
  {H.-J.}\ \bibnamefont {Choi}}, \bibinfo {author} {\bibfnamefont {J.-M.}\
  \bibnamefont {Seo}}, \bibinfo {author} {\bibfnamefont {S.-M.}\ \bibnamefont
  {Jung}}, \bibinfo {author} {\bibfnamefont {D.}~\bibnamefont {Kim}}, \bibinfo
  {author} {\bibfnamefont {F.}~\bibnamefont {Li}}, \bibinfo {author}
  {\bibfnamefont {M.~S.}\ \bibnamefont {Lah}}, \bibinfo {author} {\bibfnamefont
  {N.}~\bibnamefont {Park}}, \bibinfo {author} {\bibfnamefont {H.-J.}\
  \bibnamefont {Shin}}, \bibinfo {author} {\bibfnamefont {J.~H.}\ \bibnamefont
  {Oh}}, \ and\ \bibinfo {author} {\bibfnamefont {J.-B.}\ \bibnamefont
  {Baek}},\ }\href {\doibase 10.1073/pnas.1605318113} {\bibfield  {journal}
  {\bibinfo  {journal} {PNAS}\ }\textbf {\bibinfo {volume} {113}},\ \bibinfo
  {pages} {7414} (\bibinfo {year} {2016})}\BibitemShut {NoStop}%
\bibitem [{\citenamefont {Marzari}\ \emph {et~al.}(2012)\citenamefont
  {Marzari}, \citenamefont {Mostofi}, \citenamefont {Yates}, \citenamefont
  {Souza},\ and\ \citenamefont {Vanderbilt}}]{Marzari2012}%
  \BibitemOpen
  \bibfield  {author} {\bibinfo {author} {\bibfnamefont {N.}~\bibnamefont
  {Marzari}}, \bibinfo {author} {\bibfnamefont {A.~A.}\ \bibnamefont
  {Mostofi}}, \bibinfo {author} {\bibfnamefont {J.~R.}\ \bibnamefont {Yates}},
  \bibinfo {author} {\bibfnamefont {I.}~\bibnamefont {Souza}}, \ and\ \bibinfo
  {author} {\bibfnamefont {D.}~\bibnamefont {Vanderbilt}},\ }\href {\doibase
  10.1103/RevModPhys.84.1419} {\bibfield  {journal} {\bibinfo  {journal} {Rev.
  Mod. Phys.}\ }\textbf {\bibinfo {volume} {84}},\ \bibinfo {pages} {1419}
  (\bibinfo {year} {2012})}\BibitemShut {NoStop}%
\bibitem [{\citenamefont {Rhim}\ \emph {et~al.}(2017)\citenamefont {Rhim},
  \citenamefont {Behrends},\ and\ \citenamefont {Bardarson}}]{Rhim2017}%
  \BibitemOpen
  \bibfield  {author} {\bibinfo {author} {\bibfnamefont {J.-W.}\ \bibnamefont
  {Rhim}}, \bibinfo {author} {\bibfnamefont {J.}~\bibnamefont {Behrends}}, \
  and\ \bibinfo {author} {\bibfnamefont {J.~H.}\ \bibnamefont {Bardarson}},\
  }\href {\doibase 10.1103/PhysRevB.95.035421} {\bibfield  {journal} {\bibinfo
  {journal} {Phys. Rev. B}\ }\textbf {\bibinfo {volume} {95}},\ \bibinfo
  {pages} {035421} (\bibinfo {year} {2017})}\BibitemShut {NoStop}%
\bibitem [{\citenamefont {Fang}\ \emph {et~al.}(2012)\citenamefont {Fang},
  \citenamefont {Gilbert},\ and\ \citenamefont {Bernevig}}]{Fang2012a}%
  \BibitemOpen
  \bibfield  {author} {\bibinfo {author} {\bibfnamefont {C.}~\bibnamefont
  {Fang}}, \bibinfo {author} {\bibfnamefont {M.~J.}\ \bibnamefont {Gilbert}}, \
  and\ \bibinfo {author} {\bibfnamefont {B.~A.}\ \bibnamefont {Bernevig}},\
  }\href {\doibase 10.1103/PhysRevB.86.115112} {\bibfield  {journal} {\bibinfo
  {journal} {Phys. Rev. B}\ }\textbf {\bibinfo {volume} {86}},\ \bibinfo
  {pages} {115112} (\bibinfo {year} {2012})}\BibitemShut {NoStop}%
\bibitem [{\citenamefont {Liu}\ \emph {et~al.}()\citenamefont {Liu},
  \citenamefont {Deng},\ and\ \citenamefont {Wakabayashi}}]{Liu2018B}%
  \BibitemOpen
  \bibfield  {author} {\bibinfo {author} {\bibfnamefont {F.}~\bibnamefont
  {Liu}}, \bibinfo {author} {\bibfnamefont {H.-Y.}\ \bibnamefont {Deng}}, \
  and\ \bibinfo {author} {\bibfnamefont {K.}~\bibnamefont {Wakabayashi}},\
  }\href@noop {} {}\Eprint {http://arxiv.org/abs/arXiv:1809.10824}
  {arXiv:1809.10824} \BibitemShut {NoStop}%
\bibitem [{\citenamefont {Fukui}\ \emph {et~al.}(2005)\citenamefont {Fukui},
  \citenamefont {Hatsugai},\ and\ \citenamefont {Suzuki}}]{Fukui2005}%
  \BibitemOpen
  \bibfield  {author} {\bibinfo {author} {\bibfnamefont {T.}~\bibnamefont
  {Fukui}}, \bibinfo {author} {\bibfnamefont {Y.}~\bibnamefont {Hatsugai}}, \
  and\ \bibinfo {author} {\bibfnamefont {H.}~\bibnamefont {Suzuki}},\ }\href
  {\doibase 10.1143/JPSJ.74.1674} {\bibfield  {journal} {\bibinfo  {journal}
  {Journal of the Physical Society of Japan}\ }\textbf {\bibinfo {volume}
  {74}},\ \bibinfo {pages} {1674} (\bibinfo {year} {2005})}\BibitemShut
  {NoStop}%
\bibitem [{\citenamefont {Fujita}\ \emph {et~al.}(1996)\citenamefont {Fujita},
  \citenamefont {Wakabayashi}, \citenamefont {Nakada},\ and\ \citenamefont
  {Kusakabe}}]{Fujita1996}%
  \BibitemOpen
  \bibfield  {author} {\bibinfo {author} {\bibfnamefont {M.}~\bibnamefont
  {Fujita}}, \bibinfo {author} {\bibfnamefont {K.}~\bibnamefont {Wakabayashi}},
  \bibinfo {author} {\bibfnamefont {K.}~\bibnamefont {Nakada}}, \ and\ \bibinfo
  {author} {\bibfnamefont {K.}~\bibnamefont {Kusakabe}},\ }\href {\doibase
  10.1143/JPSJ.65.1920} {\bibfield  {journal} {\bibinfo  {journal} {J. Phys.
  Soc. Jpn.}\ }\textbf {\bibinfo {volume} {65}},\ \bibinfo {pages} {1920}
  (\bibinfo {year} {1996})}\BibitemShut {NoStop}%
\bibitem [{\citenamefont {Wakabayashi}\ \emph {et~al.}(2010)\citenamefont
  {Wakabayashi}, \citenamefont {Okada}, \citenamefont {Tomita}, \citenamefont
  {Fujimoto},\ and\ \citenamefont {Natsume}}]{Wakabayashi.jpsj.2010}%
  \BibitemOpen
  \bibfield  {author} {\bibinfo {author} {\bibfnamefont {K.}~\bibnamefont
  {Wakabayashi}}, \bibinfo {author} {\bibfnamefont {S.}~\bibnamefont {Okada}},
  \bibinfo {author} {\bibfnamefont {R.}~\bibnamefont {Tomita}}, \bibinfo
  {author} {\bibfnamefont {S.}~\bibnamefont {Fujimoto}}, \ and\ \bibinfo
  {author} {\bibfnamefont {Y.}~\bibnamefont {Natsume}},\ }\href@noop {}
  {\bibfield  {journal} {\bibinfo  {journal} {J. Phys. Soc. Jpn.}\ }\textbf
  {\bibinfo {volume} {79}},\ \bibinfo {pages} {034706} (\bibinfo {year}
  {2010})}\BibitemShut {NoStop}%
\bibitem [{\citenamefont {Soler}\ \emph {et~al.}(2002)\citenamefont {Soler},
  \citenamefont {Artacho}, \citenamefont {Gale}, \citenamefont {Garcia},
  \citenamefont {Junquera}, \citenamefont {Ordejon},\ and\ \citenamefont
  {Sanchez-Portal}}]{siesta_method}%
  \BibitemOpen
  \bibfield  {author} {\bibinfo {author} {\bibfnamefont {J.~M.}\ \bibnamefont
  {Soler}}, \bibinfo {author} {\bibfnamefont {E.}~\bibnamefont {Artacho}},
  \bibinfo {author} {\bibfnamefont {J.~D.}\ \bibnamefont {Gale}}, \bibinfo
  {author} {\bibfnamefont {A.}~\bibnamefont {Garcia}}, \bibinfo {author}
  {\bibfnamefont {J.}~\bibnamefont {Junquera}}, \bibinfo {author}
  {\bibfnamefont {P.}~\bibnamefont {Ordejon}}, \ and\ \bibinfo {author}
  {\bibfnamefont {D.}~\bibnamefont {Sanchez-Portal}},\ }\href {\doibase
  10.1088/0953-8984/14/11/302} {\bibfield  {journal} {\bibinfo  {journal} {J.
  Phys. Condens. Matter}\ }\textbf {\bibinfo {volume} {14}},\ \bibinfo {pages}
  {2745} (\bibinfo {year} {2002})}\BibitemShut {NoStop}%
\bibitem [{\citenamefont {Wakabayashi}\ \emph {et~al.}(2007)\citenamefont
  {Wakabayashi}, \citenamefont {Takane},\ and\ \citenamefont
  {Sigrist}}]{Wakabayashi2007}%
  \BibitemOpen
  \bibfield  {author} {\bibinfo {author} {\bibfnamefont {K.}~\bibnamefont
  {Wakabayashi}}, \bibinfo {author} {\bibfnamefont {Y.}~\bibnamefont {Takane}},
  \ and\ \bibinfo {author} {\bibfnamefont {M.}~\bibnamefont {Sigrist}},\ }\href
  {\doibase 10.1103/PhysRevLett.99.036601} {\bibfield  {journal} {\bibinfo
  {journal} {Phys. Rev. Lett.}\ }\textbf {\bibinfo {volume} {99}},\ \bibinfo
  {pages} {036601} (\bibinfo {year} {2007})}\BibitemShut {NoStop}%
\bibitem [{\citenamefont {Wakabayashi}\ \emph
  {et~al.}(2009{\natexlab{a}})\citenamefont {Wakabayashi}, \citenamefont
  {Takane}, \citenamefont {Yamamoto},\ and\ \citenamefont
  {Sigrist}}]{Wakabayashi2009A}%
  \BibitemOpen
  \bibfield  {author} {\bibinfo {author} {\bibfnamefont {K.}~\bibnamefont
  {Wakabayashi}}, \bibinfo {author} {\bibfnamefont {Y.}~\bibnamefont {Takane}},
  \bibinfo {author} {\bibfnamefont {M.}~\bibnamefont {Yamamoto}}, \ and\
  \bibinfo {author} {\bibfnamefont {M.}~\bibnamefont {Sigrist}},\ }\href
  {http://stacks.iop.org/1367-2630/11/i=9/a=095016} {\bibfield  {journal}
  {\bibinfo  {journal} {New Journal of Physics}\ }\textbf {\bibinfo {volume}
  {11}},\ \bibinfo {pages} {095016} (\bibinfo {year}
  {2009}{\natexlab{a}})}\BibitemShut {NoStop}%
\bibitem [{\citenamefont {Wakabayashi}\ \emph
  {et~al.}(2009{\natexlab{b}})\citenamefont {Wakabayashi}, \citenamefont
  {Takane}, \citenamefont {Yamamoto},\ and\ \citenamefont
  {Sigrist}}]{Wakabayashi2009B}%
  \BibitemOpen
  \bibfield  {author} {\bibinfo {author} {\bibfnamefont {K.}~\bibnamefont
  {Wakabayashi}}, \bibinfo {author} {\bibfnamefont {Y.}~\bibnamefont {Takane}},
  \bibinfo {author} {\bibfnamefont {M.}~\bibnamefont {Yamamoto}}, \ and\
  \bibinfo {author} {\bibfnamefont {M.}~\bibnamefont {Sigrist}},\ }\href
  {\doibase https://doi.org/10.1016/j.carbon.2008.09.040} {\bibfield  {journal}
  {\bibinfo  {journal} {Carbon}\ }\textbf {\bibinfo {volume} {47}},\ \bibinfo
  {pages} {124 } (\bibinfo {year} {2009}{\natexlab{b}})}\BibitemShut {NoStop}%
\bibitem [{\citenamefont {Dutta}\ and\ \citenamefont
  {Wakabayashi}(2012)}]{dutta_2012}%
  \BibitemOpen
  \bibfield  {author} {\bibinfo {author} {\bibfnamefont {S.}~\bibnamefont
  {Dutta}}\ and\ \bibinfo {author} {\bibfnamefont {K.}~\bibnamefont
  {Wakabayashi}},\ }\href {\doibase 10.1039/C2JM34881K} {\bibfield  {journal}
  {\bibinfo  {journal} {J. Mater. Chem.}\ }\textbf {\bibinfo {volume} {22}},\
  \bibinfo {pages} {20881} (\bibinfo {year} {2012})}\BibitemShut {NoStop}%
\bibitem [{\citenamefont {Dutta}\ and\ \citenamefont
  {Wakabayashi}(2013)}]{dutta_2013}%
  \BibitemOpen
  \bibfield  {author} {\bibinfo {author} {\bibfnamefont {S.}~\bibnamefont
  {Dutta}}\ and\ \bibinfo {author} {\bibfnamefont {K.}~\bibnamefont
  {Wakabayashi}},\ }\href {\doibase 10.1039/C3TC31136H} {\bibfield  {journal}
  {\bibinfo  {journal} {J. Mater. Chem. C}\ }\textbf {\bibinfo {volume} {1}},\
  \bibinfo {pages} {4854} (\bibinfo {year} {2013})}\BibitemShut {NoStop}%
\end{thebibliography}%

\end{document}